\documentclass[]{article}

\usepackage{amsmath}
\usepackage{bm}
\usepackage[dvips,xdvi]{graphicx}
\newcommand{\citen}[1]{\cite{#1}}

\begin{document}


\title{\bf Calculation of the Micellar Structure of Polymer Surfactant Based 
on the Density Functional Theory}
\author{ Takashi Uneyama${}^{1}$ and Masao Doi${}^{2}$ \\
\\
${}^{1}$ Department of Physics, \\
Graduate School of Science, Kyoto University \\
Sakyo-ku, Kyoto 606-8502, JAPAN \\
${}^{2}$ Department of Applied Physics, \\
Graduate School of Engineering, The University of Tokyo, CREST JST \\
Hongo, Tokyo 113-8656, JAPAN}
\date{}
\maketitle

%
%

\begin{abstract}
Amphiphilic block copolymer solutions form various micellar structures
including micelles and vesicles. We applied the density functional
theory for block copolymers which we have proposed to amphiphilic block
copolymer systems. The 3 dimensional simulation for AB diblock copolymer
solutions and AB diblock copolymer / A homopolymer blends has been done
and it is shown that the spherical micelles, cylindrical micelles and
spherical vesicles are formed.  It is also shown that the phase diagram
for AB diblock copolymer / A homopolymer blends qualitatively agrees
with the phase diagram obtained by the experiment.
\end{abstract}


%
%

\section{Introduction}

Block copolymers dissolved in solvent or other polymers assemble
spontaneously to form various self organized structures such as spheres,
rods and vesicles \cite{Choucair-Eisenberg-2003}. What structure is
formed is a question very important in colloid science, but difficult to
answer since the structure depends on many parameters such as the block
ratio, the affinity between component units, the polymerization index and
the volume fraction of the block copolymers. Attempt has been made to
predict the micellar structure by computer simulation. Since the
process of self-organization involves many molecules and is very slow,
atomistic molecular dynamics
simulation cannot be used for this purpose. 

Various coarse grained
models have been used to study the formation of the micellar
structure. Larson et al. \cite{Larson-1988,Larson-1989,Larson-1992} used
a lattice model to show that amphiphilic molecules actually form micellar
structure . Bernardes \cite{Bernardes-1996} studied the vesicle
formation by the Larson type lattice model. Yamamoto \cite{Yamamoto-Maruyama-Hyodo-2002}
 demonstrated that the DPD
(dissipative particle dynamics)  \cite{Groot-Warren-1997} can show the formation and various
structural change of vesicles. Similar attempt has been done by Noguchi \cite{Noguchi-Takasu-2001,Noguchi-2002}
et al. for small molecule surfactant solutions. They studied formation
and deformation of surfactant vesicles by Brownian Dynamics (BD).

In these models it is not
straightforward to connect the coarse grained model with the underlying
atomistic model. (Some parameters can be associated with the mean field
parameter. For example, the interaction parameters used in DPD can be mapped onto
 the Flory-Huggins $\chi$ parameter
 \cite{Groot-Warren-1997,Maiti-McGrother-2004}.
But this is not always successful, and needs more
validation.)

Other approach which takes into account of the molecular characteristics
is the dynamic mean field theory which is based on the self 
consistent field (SCF) theory \cite{Drolet-Fredrickson-1999,Fredrickson-Ganesan-Drolet-2002,Fraaije-1993,Kawakatsu-book}.
In this approach, it is possible
to determine the model parameters from the atomistic models for certain
classes of polymers. 

For periodic systems, computationally efficient and fast SCF method
has been developed \cite{Matsen-Schick-1994}
which uses eigen functions. The method
has been applied for AB diblock copolymer / A homopolymer 
blends which forms micelles \cite{Matsen-1995}. However the
method is not best suited to study the micelles or vesicles
which does not form periodic structures.  

For non-periodic systems, the real-space SCF simulation method
has been frequently used. The method is based on the real-space
calculation of the path 
integral and needs large CPU power and memory. 
Although considerable development has been achieved due to the improvement of
algorithms \cite{Drolet-Fredrickson-1999,Fredrickson-Ganesan-Drolet-2002}
and in models \cite{Fraaije-1993}, it is still computationally demanding
for large systems.

Some works has been done for micellar systems using the 
real-space SCF simulation. For example,
Fraaije et al. \cite{vamVlimmeren-Maurits-Zvelindovsky-Sevink-Fraaije-1999} 
and Lam et al. \cite{Lam-GoldbeckWood-2003} have studied rather dense block
copolymer solution with the mean field parameters calculated from the
experimental data and shown that they form the micellar structure.
Fraiije and Sevink \cite{Fraaije-Sevink-2003} also demonstrated various
intriguing self organized structure formed in the droplet of
concentrated solution of block copolymers in equilibrium with the outer
solvent phase.

While these applications of SCF have achieved successful
results, the dilute micellar systems have not been 
studied yet except for a few works.
He et al. \cite{He-Liang-Huang-Pan-2004}. performed 2 dimensional SCF
calculation for the dilute AB diblock copolymer solution
(AB diblock copolymer / S solvent mixtures) and showed that the AB
diblock copolymer actually forms 2 dimensional vesicles. Though
this is an important achievement, the 2 dimensional simulation
poses problems in the interpretation of the results. The spherical
vesicles are 3 dimensional objects and cannot be handled by 2
dimensional simulation. At present, the 3 dimensional
real-space SCF simulation for micellar systems 
costs too much CPU power and memory and difficult to perform.

The computational cost of the dynamic mean field theory can be reduced
drastically if one knows the free energy of the system as the functional
of the density distribution of each component. The crucial point of this
approach is the expression of the density functional. Earlier works
\cite{Kawakatsu-1994,Ohta-Ito-1995} used heuristic arguments for the
expression, and was difficult to generalize. In the previous paper
\cite{Uneyama-Doi-2005}, we have proposed a
general expression for the density
functional which involves the same parameters appearing in the SCF
theory. We have shown some preliminary results
which shows the formation of the self organized structure for the
mixture of block copolymers and homopolymers. 

In this paper, we shall show that our free energy functional can be
applied and is effective for the study of 3 dimensional micellar systems.
We shall show that micelles and vesicles are formed from the
homogeneous state.  (As far as we know, this is the first 3 dimensional
simulation of the vesicle formation from the homogeneous state using 
the continuous field model.) We shall discuss the relationship
between the micellar structure and the molecular parameters (volume
fractions, $\chi$ parameters, the block ratio).  We shall
also discuss the comparison between the present simulation and 
experiments and the SCF simulation by He et
al. \cite{He-Liang-Huang-Pan-2004}.

%
%

\section{Free Energy Functional Model}
In this paper, we discuss two systems: one is the
AB diblock copolymer solution (the solvent being a small
molecule of C), and the other is a polymer blend, a mixture of 
AB diblock copolymer and A homopolymer. 
In both systems the block copolymers act as the
surfactant and form the self organizing, micellar
structures.

In the density functional theory, the free energy $F$ of the system is
written as a functional of the density distribution of the
monomers belonging to various subchains in the system.  The subchains
are indexed by $(p,i)$, where $p$ and $i$ refer to the polymer, and the
subchain in the polymer.  In the present problem, the system consists of
three subchains (AB,A), (AB,B) and (S,S), each of which
stands for the A block of the AB diblock copolymer, 
the B block of the AB diblock copolymer, and 
the matrix, which is the C solvent in the case of solution, and
A homopolymer in the case of polymer blend.

Let $\phi_{pi}(\bm{r})$ be the volume fraction of the monomers belonging
to the $(p,i)$ subchain at point $\bm{r}$. 
We use the following free energy functional model \cite{Uneyama-Doi-2005}.
\begin{equation}
 \label{freeenergy}
 \begin{split}
  F \left[\{\phi_{pi}(\bm{r})\}\right] = &
  \sum_{p,ij}  \int d\bm{r} d\bm{r}' \, 2 \sqrt{f_{p i} f_{pj}} A_{p,ij}
  \mathcal{G}(\bm{r} - \bm{r}') \sqrt{\phi_{pi}(\bm{r}) \phi_{pj}(\bm{r}')} \\
  & + \sum_{pi}  \int d\bm{r} \, f_{pi} C_{p,ii} 
  \phi_{pi}(\bm{r}) \ln \phi_{pi}(\bm{r}) \\
  & + \sum_{p,i \neq j} \int d\bm{r} \, 2 \sqrt{f_{pi} f_{pj}} C_{p,ij} 
  \sqrt{\phi_{pi}(\bm{r}) \phi_{pj}(\bm{r})} \\
  & + \sum_{pi} \int d\bm{r} \, \frac{b^{2}}{24 \phi_{pi}(\bm{r})}
  \left| \nabla \phi_{pi}(\bm{r}) \right|^{2} \\
  & + \sum_{pi,qj} \int d\bm{r} \, \frac{\chi_{pi,qj}}{2} \phi_{pi}(\bm{r}) \phi_{qj}(\bm{r})
 \end{split}
\end{equation}
where $\mathcal{G}(\bm{r} - \bm{r}')$ is the Green function which
satisfies $- \nabla^{2} \mathcal{G}(\bm{r} - \bm{r}') = \delta(\bm{r} -
\bm{r}')$, $f_{pi}$ is the block ratio of the $i$-th subchain of $p$-th
polymer, $b$ is the effective bond length and $\chi_{pi,qj}$ is the
Flory-Huggins $\chi$ parameter. $A_{p,ij}$ and $C_{p,ij}$ are determined
from $b$, $f_{pi}$, the degree of polymerization $N_{p}$ and the structure of block copolymer 
(see Ref.~\citen{Uneyama-Doi-2005} for detail).

The free energy functional eq~\eqref{freeenergy} reduces to the following
form for the AB diblock copolymer / S solvent blends.
\begin{equation}
 \label{freeenergy_diblocksolution}
 \begin{split}
  F \left[\phi_{A}(\bm{r}), \phi_{B}(\bm{r}), \phi_{S}(\bm{r}) \right] = &
  \sum_{\substack{i,j \\ (= A,B)}}  \int d\bm{r} d\bm{r}' \, 2 \sqrt{f_{i} f_{j}} A_{ij}
  \mathcal{G}(\bm{r} - \bm{r}') \sqrt{\phi_{i}(\bm{r}) \phi_{j}(\bm{r}')} \\
  & + \int d\bm{r} \, \left[f_{A} C_{AA} \phi_{A}(\bm{r}) \ln \phi_{A}(\bm{r}) 
  + f_{B} C_{BB} \phi_{B}(\bm{r}) \ln \phi_{B}(\bm{r}) \right] \\
  & + \int d\bm{r} \, 4 \sqrt{f_{A} f_{B}} C_{AB} 
  \sqrt{\phi_{A}(\bm{r}) \phi_{B}(\bm{r})} \\
  & + \int d\bm{r} \, \left[ \frac{b^{2}}{24 \phi_{A}(\bm{r})} \left| \nabla \phi_{A}(\bm{r}) \right|^{2}
  + \frac{b^{2}}{24 \phi_{B}(\bm{r})} \left| \nabla \phi_{B}(\bm{r}) \right|^{2} \right] \\
  & + \int d\bm{r} \, \frac{1}{N_{S}} \phi_{S}(\bm{r}) \ln \phi_{S}(\bm{r}) \\
  & + \int d\bm{r} \, \frac{b^{2}}{24 \phi_{S}(\bm{r})} \left| \nabla \phi_{S}(\bm{r}) \right|^{2} \\
  & + \sum_{\substack{i,j \\ (= A,B,S)}} \int d\bm{r} \, \frac{\chi_{i,j}}{2} \phi_{i}(\bm{r}) \phi_{j}(\bm{r})
 \end{split}
\end{equation}
Here we dropped the index $p$, which specify the polymer species, for
simplicity. Thus $\phi_{A}(\bm{r})$, $\phi_{B}(\bm{r})$ and 
$\phi_{S}(\bm{r})$ stands for the density of the A component of
the block copolymer, the density of the B component of the block copolymer
and the density of the matrix (C solvent in the case of
polymer solution and A homopolymer in the case of polymer blend).
$A_{ij}$ and $C_{ij}$ for diblock copolymer ($i,j = A,B$) 
is given in Appendix \ref{form_of_matrices_a_c}.

It is convenient to introduce the new order parameter
(the $\psi$-field) defined via
the following equation \cite{Uneyama-Doi-2005};
\begin{equation}
 \label{psi_definition}
 \psi_{i}(\bm{r}) \equiv \sqrt{\phi_{i}(\bm{r})}
\end{equation}
With this order parameter, eq~\eqref{freeenergy_diblocksolution} is
rewritten as
\begin{equation}
 \label{freeenergy_diblocksolution_psi}
 \begin{split}
  F \left[\psi_{A}(\bm{r}), \psi_{B}(\bm{r}), \psi_{S}(\bm{r}) \right] = &
  \sum_{\substack{i,j \\ (= A,B)}}  \int d\bm{r} d\bm{r}' \, 2 \sqrt{f_{i} f_{j}} A_{ij}
  \mathcal{G}(\bm{r} - \bm{r}') \psi_{i}(\bm{r}) \psi_{j}(\bm{r}') \\
  & + \int d\bm{r} \, \left[2 f_{A} C_{AA} \psi^{2}_{A}(\bm{r}) \ln \psi_{A}(\bm{r}) 
  + 2 f_{B} C_{BB} \psi^{2}_{B}(\bm{r}) \ln \psi_{B}(\bm{r}) \right] \\
  & + \int d\bm{r} \, 4 \sqrt{f_{A} f_{B}} C_{AB}
  \psi_{A}(\bm{r}) \psi_{B}(\bm{r}) \\
  & + \int d\bm{r} \, \left[ \frac{b^{2}}{6} \left| \nabla \psi_{A}(\bm{r}) \right|^{2}
  + \frac{b^{2}}{6} \left| \nabla \psi_{B}(\bm{r}) \right|^{2} \right] \\
  & + \int d\bm{r} \, \frac{2}{N_{S}} \psi^{2}_{S}(\bm{r}) \ln \psi_{S}(\bm{r}) \\
  & + \int d\bm{r} \, \frac{b^{2}}{6} \left| \nabla \psi_{S}(\bm{r}) \right|^{2} \\
  & + \sum_{\substack{i,j \\ (= A,B,S)}} \int d\bm{r} \, \frac{\chi_{i,j}}{2} \psi^{2}_{i}(\bm{r}) \psi^{2}_{j}(\bm{r})
 \end{split}
\end{equation}

%
%

\section{Simulation}

We performed the simulation to get the equilibrium structure of the
diblock copolymer solution. To get the equilibrium structure
numerically, we minimized the free energy functional
eq~\eqref{freeenergy_diblocksolution_psi} by using the algorithm shown
in this section.

\subsection{Constraints}

The minimization of the free energy was done under 
two constraints each representing the conservation condition and
the incompressible condition.
\begin{align}
 \label{conservation}
 & \int d\bm{r} \, \psi^{2}_{i}(\bm{r}) = V \bar{\phi}_{i}, \qquad
  (i = A,B,S) \\
 \label{incompressible}
 & \sum_{i (= A,B,S)} \psi^{2}_{i}(\bm{r}) = 1
\end{align}
where $V$ is the volume of the system and
$\bar{\phi}_{i}$ is the spatial average of $\phi_{i}(\bm{r})$.
 $\bar{\phi}_{S}$ is the average volume
fraction of solvent and $\bar{\phi}_{A}, \bar{\phi}_{B}$ are given by the
following equations.
\begin{align}
 \bar{\phi}_{A} & = f_{A} \bar{\phi}_{AB} \\
 \bar{\phi}_{B} & = f_{B} \bar{\phi}_{AB}
\end{align}
where $\bar{\phi}_{AB}$ is the average volume fraction of
the AB diblock copolymer. $\bar{\phi}_{AB}$ and $\bar{\phi}_{S}$ 
satisfy the following equation.
\begin{equation}
 \bar{\phi}_{AB} + \bar{\phi}_{S} = 1
\end{equation}
Constraints eqs~\eqref{conservation}, \eqref{incompressible}
give the following terms for the free energy functional \eqref{freeenergy_diblocksolution_psi}.
\begin{equation}
 \label{freeenergy_constraints}
  F_{\text{constraints}} \left[\psi_{A}(\bm{r}), \psi_{B}(\bm{r}), \psi_{S}(\bm{r}) \right] = 
  \sum_{i} \int d\bm{r} \, \frac{1}{2} \left[ \lambda_{i} + \kappa(\bm{r}) \right]
  \left[\psi_{i}^{2}(\bm{r}) - \bar{\phi}_{i} \right]
\end{equation}
where $\lambda_{i}$ and $\kappa(\bm{r})$ are the Lagrangian multipliers 
which correspond to the conservation and incompressible conditions, respectively.

\subsection{Numerical Scheme}

The free energy was minimized by evolving the $\psi$-field iteratively by using
the steepest-descent method.
\begin{equation}
 \label{steepest_descent}
 \psi^{\text{(new)}}_{i}(\bm{r}) = \psi_{i}(\bm{r}) 
 - \omega \left[ \mu_{i}(\bm{r}) + \lambda_{i} \psi_{i}(\bm{r}) + \kappa(\bm{r}) \psi_{i}(\bm{r}) \right]
\end{equation}
where $\omega$ is positive constant and $\mu_{i}(\bm{r})$ is the
chemical potential defined as
\begin{equation}
 \label{chemicalpotential}
  \mu_{i}(\bm{r}) \equiv \frac{\delta F \left[ \psi_{A}(\bm{r}), \psi_{B}(\bm{r}), \psi_{S}(\bm{r}) \right]}{\delta \psi_{i}(\bm{r})} 
\end{equation}
By substituting eq~\eqref{freeenergy_diblocksolution_psi}
into eq~\eqref{chemicalpotential}, we get the explicit form of the
chemical potential (See Appendix \ref{chemical_potential} for detail).

We employed the time-splitting method and the ADI method
\cite{NumericalRecipes} to update $\psi^{\text{(new)}}(\bm{r})$ from
$\psi(\bm{r})$ by eq~\eqref{steepest_descent}.
\begin{align}
 \label{timesplitting_x}
 \psi^{(1)}_{i}(\bm{r}) & = \psi_{i}(\bm{r}) 
 - \omega^{(1)} \left\{ \left[\mu_{i}(\bm{r}) 
 + \frac{b^{2}}{3} \frac{\partial^{2} \psi_{i}(\bm{r})}{\partial x^{2}} \right] 
 - \frac{b^{2}}{3} \frac{\partial^{2} \psi^{(1)}(\bm{r})}{\partial x^{2}} \right\} \\
 \label{timesplitting_y}
 \psi^{(2)}_{i}(\bm{r}) & = \psi^{(1)}_{i}(\bm{r}) 
 - \omega^{(1)} \left\{ \left[\mu^{(1)}_{i}(\bm{r})
 + \frac{b^{2}}{3} \frac{\partial^{2} \psi^{(1)}_{i}(\bm{r})}{\partial y^{2}} \right] 
 - \frac{b^{2}}{3} \frac{\partial^{2} \psi^{(2)}(\bm{r})}{\partial y^{2}} \right\} \\
 \label{timesplitting_z}
 \psi^{(3)}_{i}(\bm{r}) & = \psi^{(2)}_{i}(\bm{r})  
 - \omega^{(1)} \left\{ \left[\mu^{(2)}_{i}(\bm{r})
 + \frac{b^{2}}{3} \frac{\partial^{2} \psi^{(2)}_{i}(\bm{r})}{\partial z^{2}} \right] 
 - \frac{b^{2}}{3} \frac{\partial^{2} \psi^{(3)}(\bm{r})}{\partial z^{2}} \right\} \\
 \label{timesplitting_incompressible}
 \psi^{(4)}_{i}(\bm{r}) & = \psi^{(3)}_{i}(\bm{r}) 
 - \omega^{(2)} \kappa(\bm{r}) \psi^{(3)}_{i}(\bm{r}) \\
 \label{timesplitting_conservation}
 \psi^{\text{(new)}}_{i}(\bm{r}) & = \psi^{(4)}_{i}(\bm{r})
 - \omega^{(3)} \lambda_{i} \psi^{(4)}
\end{align}
where $\omega^{(n)}$ is positive constant and
$\mu^{(n)}_{i}$ is the chemical potential calculated from $\psi^{(n)}_{i}(\bm{r})$.
The advantage of using the ADI method is that 
we can choose a large value for $\omega$ compared with one
used in the explicit method which we employed in our previous work
\cite{Uneyama-Doi-2005}.
The evolution with eq~\eqref{timesplitting_conservation} can be regarded
as rescaling of the $\psi$-field to satisfy the conservation law (this is
analogy to the wave function in quantum mechanics; we normalize the wave
function to get probability density). Thus
eq~\eqref{timesplitting_conservation} was modified as follows.
\begin{equation}
 \label{timesplitting_conservation2}
 \psi^{\text{(new)}}_{i}(\bm{r}) = \frac{V \bar{\phi}_{i}}{\displaystyle \int d\bm{r} \, \left[\psi^{(4)}_{i}(\bm{r}) \right]^{2}} \, \psi^{(4)}_{i}(\bm{r})
\end{equation}
We also evolved $\kappa(\bm{r})$ as well.
\begin{equation}
 \label{timesplitting_kappa}
 \kappa^{\text{(new)}}(\bm{r}) = \kappa(\bm{r}) - \omega^{(4)}
 \left[ \psi^{2}_{i}(\bm{r}) - \bar{\phi}_{i} \right]
\end{equation}

Eqs~\eqref{timesplitting_x}-\eqref{timesplitting_incompressible},\eqref{timesplitting_conservation2}
and \eqref{timesplitting_kappa} were solved iteratively to get
the steady state structure.
All simulations were started from the nearly homogeneous state with small
Gaussian white noise and the periodic boundary condition was used.
 The evolution was conducted until the $\psi$ field
do not change anymore. This structure is not necessarily the
equilibrium structure: in most cases it is a 
metastable structure. As our purpose is to search various metastable
structures, we did not pursue to find the real minimum of the free energy.

The numerical scheme described above produces the
equilibrium (or metastable) structure efficiently. For example,
we can get 3 dimensional AB diblock copolymer vesicles (lattice points:
$128 \times 128 \times 128$, detail parameters are shown in the following
section) starting from the homogeneous state 
in about 26 hours on a 3.0GHz Pentium 4 PC.

For the systems with small volume fraction of the block copolymer and/or
small $\chi$ parameter, spontaneous formation of the self-organized
structure from the homogeneous state was not always successful.
It is difficult to get the micellar structure by our scheme especially
for these systems.
Thus we used the following procedure. 

We started the simulations from the nearly homogeneous state 
by using rather large $\chi$ parameter (for example
$\chi_{AB} = 1$) and/or large volume fraction (for example
$\bar{\phi}_{AB} = 0.1$). This gives a state which has a large 
density fluctuation after several hundreds iteration steps.
We then switched $\chi_{AB}$ and $\bar{\phi}_{AB}$ to the target value
and continued the simulation for several thousand iteration steps (about $2000$ to $4000$ in
this work) to get the equilibrium state.  Though arbitrary and
artificial it may look, 
such procedure was needed in order to get the phase separated
structure. In fact, without the procedure, we could not
reproduce the phase separation in the region where 
the $\chi$ parameter or the volume fraction are small.

%
%

\section{Result}

\subsection{AB Diblock Copolymer Solutions (AB Diblock Copolymer / C Solvent)}
First we show the result of the simulation for the AB diblock copolymer
solutions. We set the A monomer is solvent-philic and the B monomer is
solvent-phobic (i.e. the $\chi$ parameter between the A and C monomers is small
or negative, and that between the B and C monomers is large).

\begin{figure}[p!]
 \centering
 {}
 {\includegraphics[width=0.45\linewidth,clip]{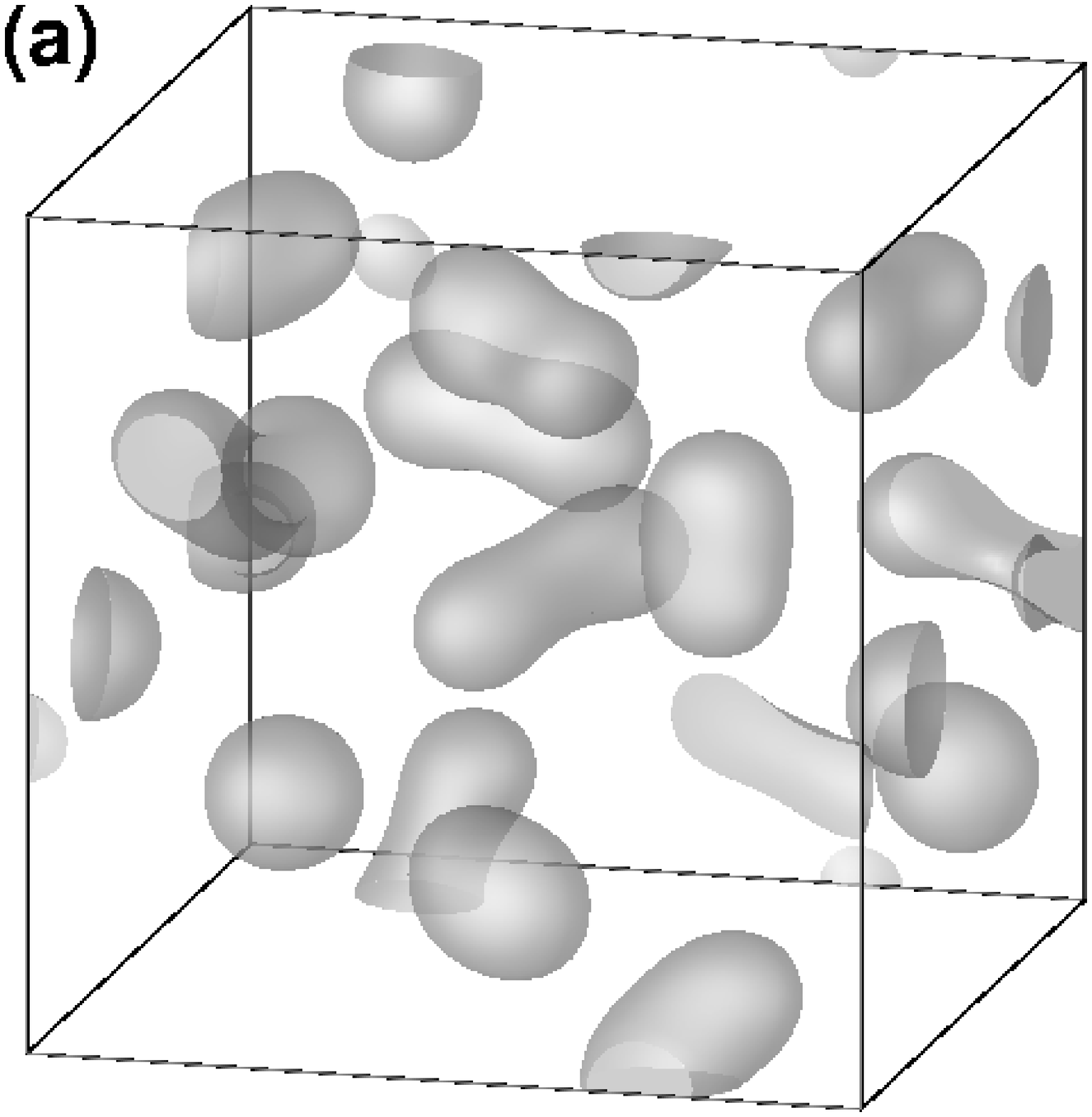}}
 \hspace{0.025\linewidth}
 {\includegraphics[width=0.45\linewidth,clip]{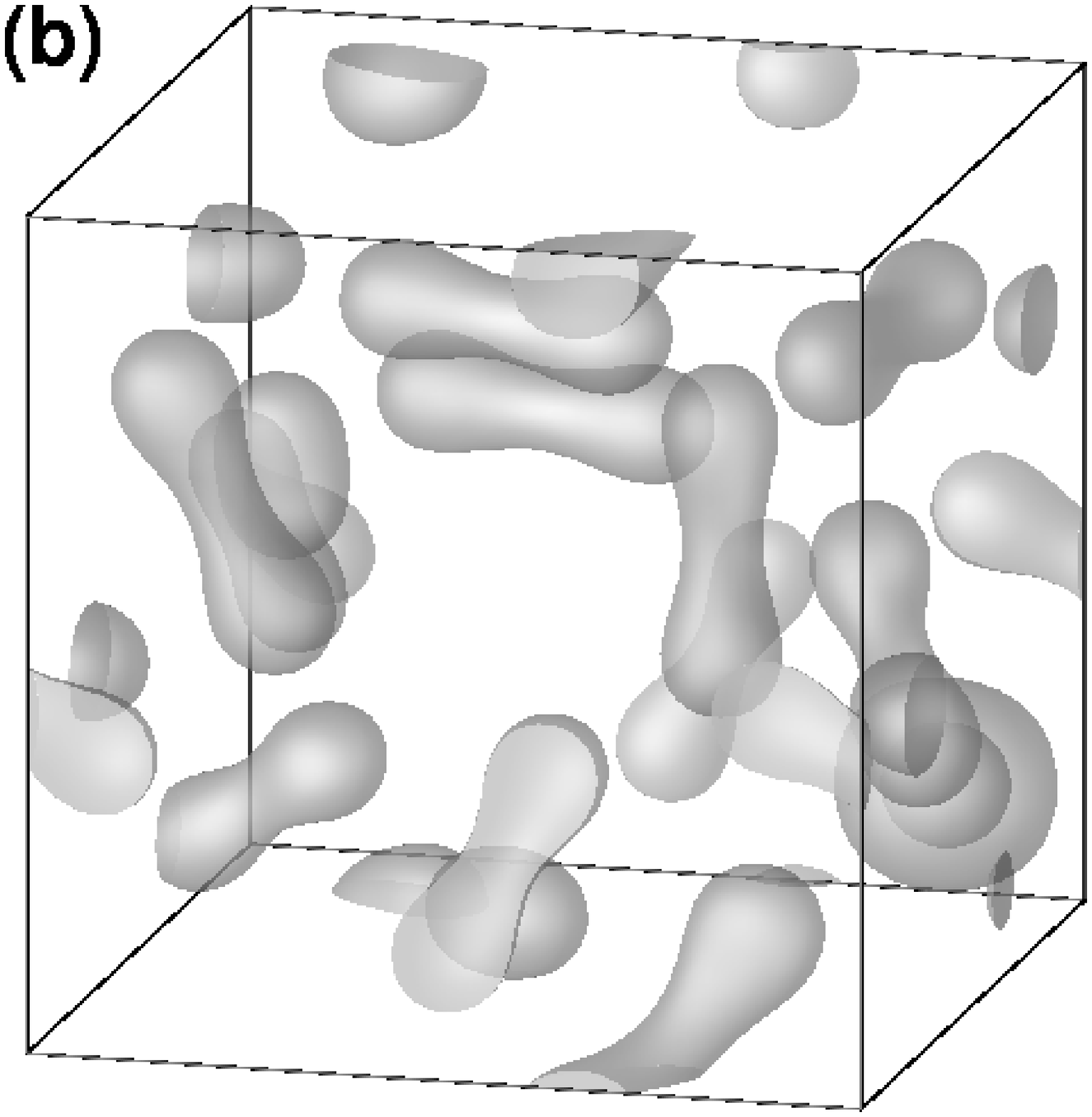}}
 \vspace{0.035\linewidth} {}\
 {\includegraphics[width=0.45\linewidth,clip]{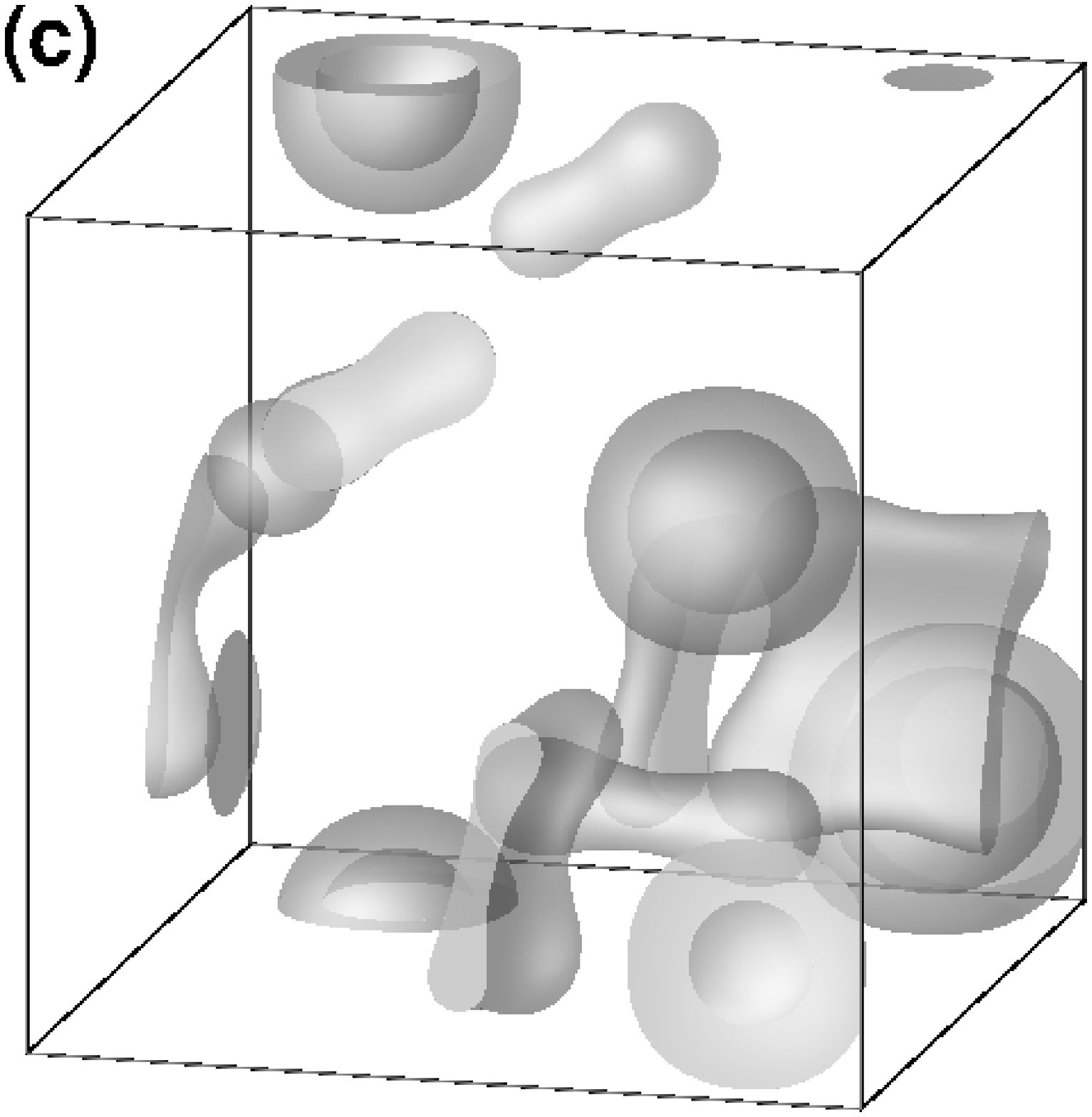}}
 \hspace{0.025\linewidth}
 \hspace{0.45\linewidth}
 \caption{Micellar structures of the AB diblock copolymer solutions (AB
 diblock copolymer / C solvent blends) with different $\chi$ parameters.
 The system size is $48 b \times 48 b \times 48 b$ ($96 \times 96 \times 96$ lattice points).
 Parameters are set
 to $N_{AB} = 20, N_{C} = 1, \bar{\phi}_{AB} = 0.1, \bar{\phi}_{C} = 0.9, \chi_{AB} = 1, \chi_{BC} = 1.75, f_{A} = 1/3, f_{B} = 2/3$.
 (a) $\chi_{AC} = 0.5$,
 (b) $\chi_{AC} = 0$,
 (c) $\chi_{AC} = -0.175$.
 The gray surfaces are isodensity surfaces for $\phi_{B}(\bm{r}) = 0.5$.}
 \label{solution}
\end{figure}

Figure \ref{solution} shows the result of the simulation.
Here the surfactant is a medium sized block copolymer ($N_{AB}=20$)
with short solvent-philic part ($f_{A} = 1/3$) and long
solvent-phobic part ($f_{B} = 2/3$). The other parameters are 
$N_{C} = 1, \bar{\phi}_{AB} = 0.1, \bar{\phi}_{C} = 0.9, \chi_{AB} = 1, \chi_{BC} = 1.75$, and
$\chi_{AC}$ are taken to be 0.5 (a), 0 (b) and -0.175 (c). 
The system size is $48b \times 48b \times 48b$ 
and the number of lattice points is $96 \times 96 \times 96$.

The figure demonstrates that the density functional method gives self
organized structures spontaneously starting from
a nearly homogeneous initial state.
Various micellar structures -- spherical micelles, cylindrical micelles, open
bilayers (not closed membrane structures) and vesicles are observed.
Notice that the micellar structure is not the same: there is a distribution
of micellar size as well as the micellar structure.  We believe that 
real micellar system will also have such distribution.

The morphological change shown in Figure \ref{solution} can be qualitatively
understood as follows.  As $\chi_{AC}$ decreases, the affinity
between the solvent and the solvent-philic part increases.  Therefore
the interfacial area between the micelles and solvent tends to
increase.  As a result, the micellar structure changes from spheres
to cylinders and then to vesicles.

\begin{figure}[p!]
 \centering
 {}
 {\includegraphics[width=0.375\linewidth,clip]{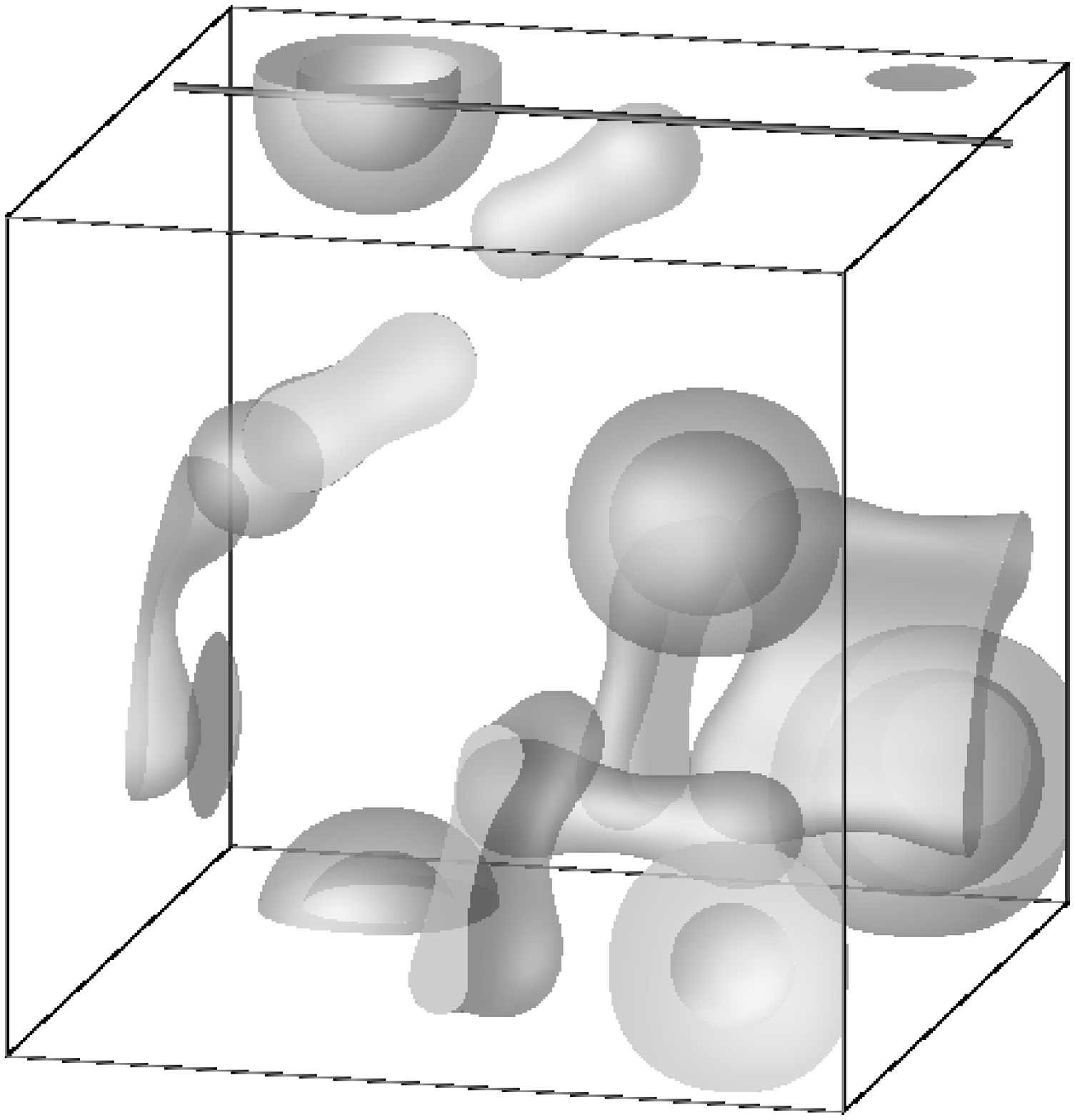}}
 \hspace{0.025\linewidth}
 {\includegraphics[width=0.525\linewidth,clip]{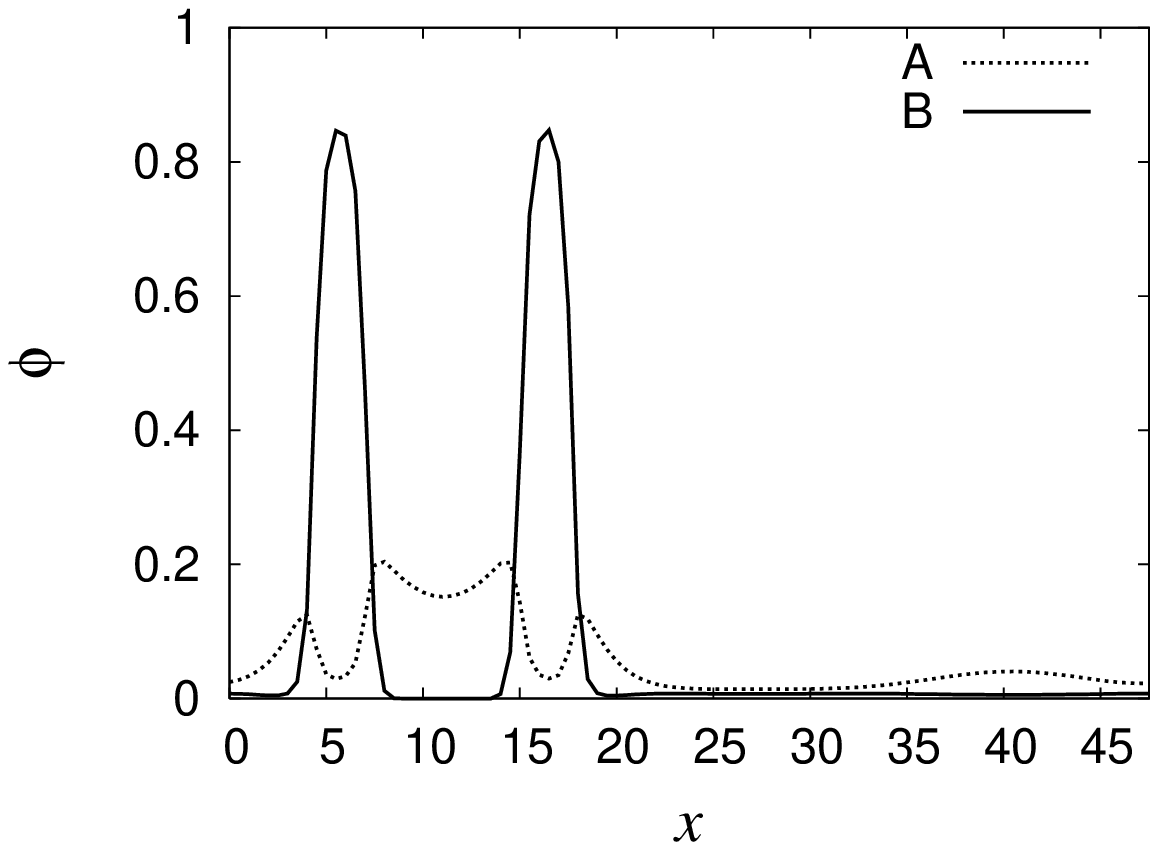}}
 \caption{Density profile of an AB diblock copolymer vesicle.
 The parameters used are the same as Figure \ref{solution}(c).
 The 1 dimensional density profile (right) is
 the profile along the black line in the 3 dimensional isosurface data (left).
 }
 \label{density_profile}
\end{figure}

Figure \ref{density_profile} shows the 1 dimensional density profile of 
Figure \ref{solution}(c). It is observed that 
AB diblock copolymer is strongly localized at the micellar 
structure (the vesicle) and the density of unimers are quite small.

\begin{figure}[p!]
 \centering
 {}
 {\includegraphics[width=0.45\linewidth,clip]{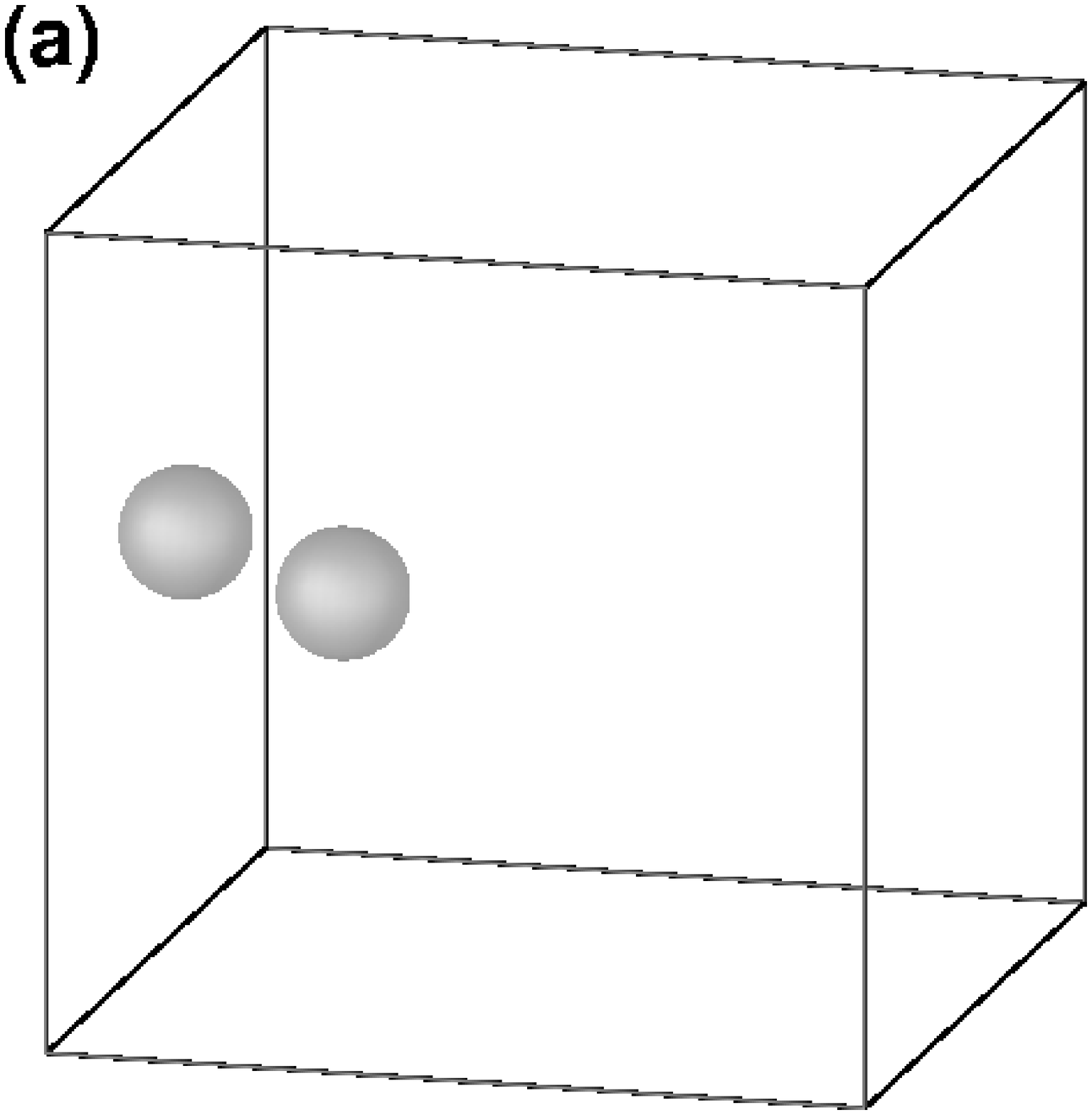}}
 \hspace{0.025\linewidth}
 {\includegraphics[width=0.45\linewidth,clip]{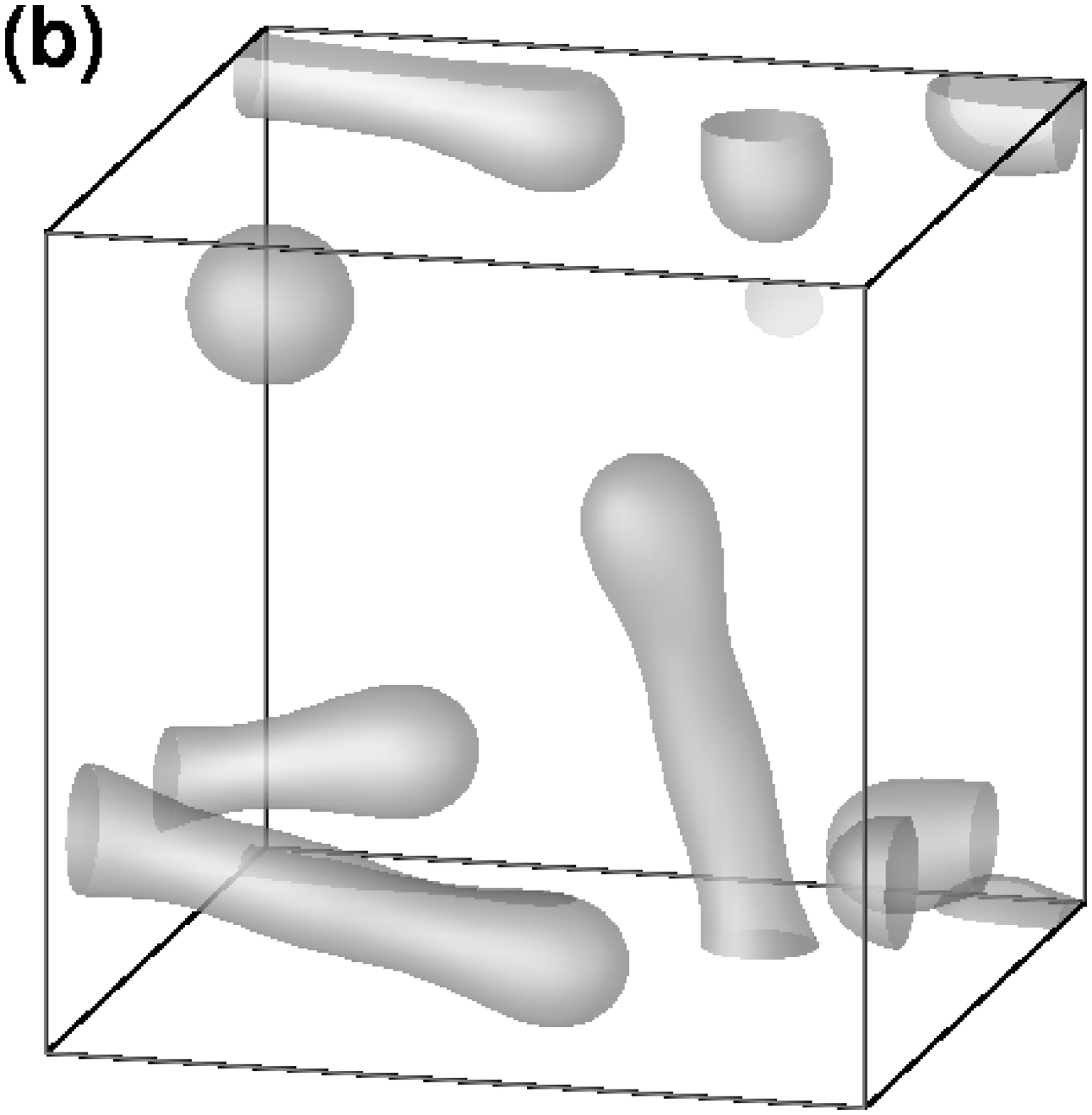}}
 \vspace{0.035\linewidth} {}
 {\includegraphics[width=0.45\linewidth,clip]{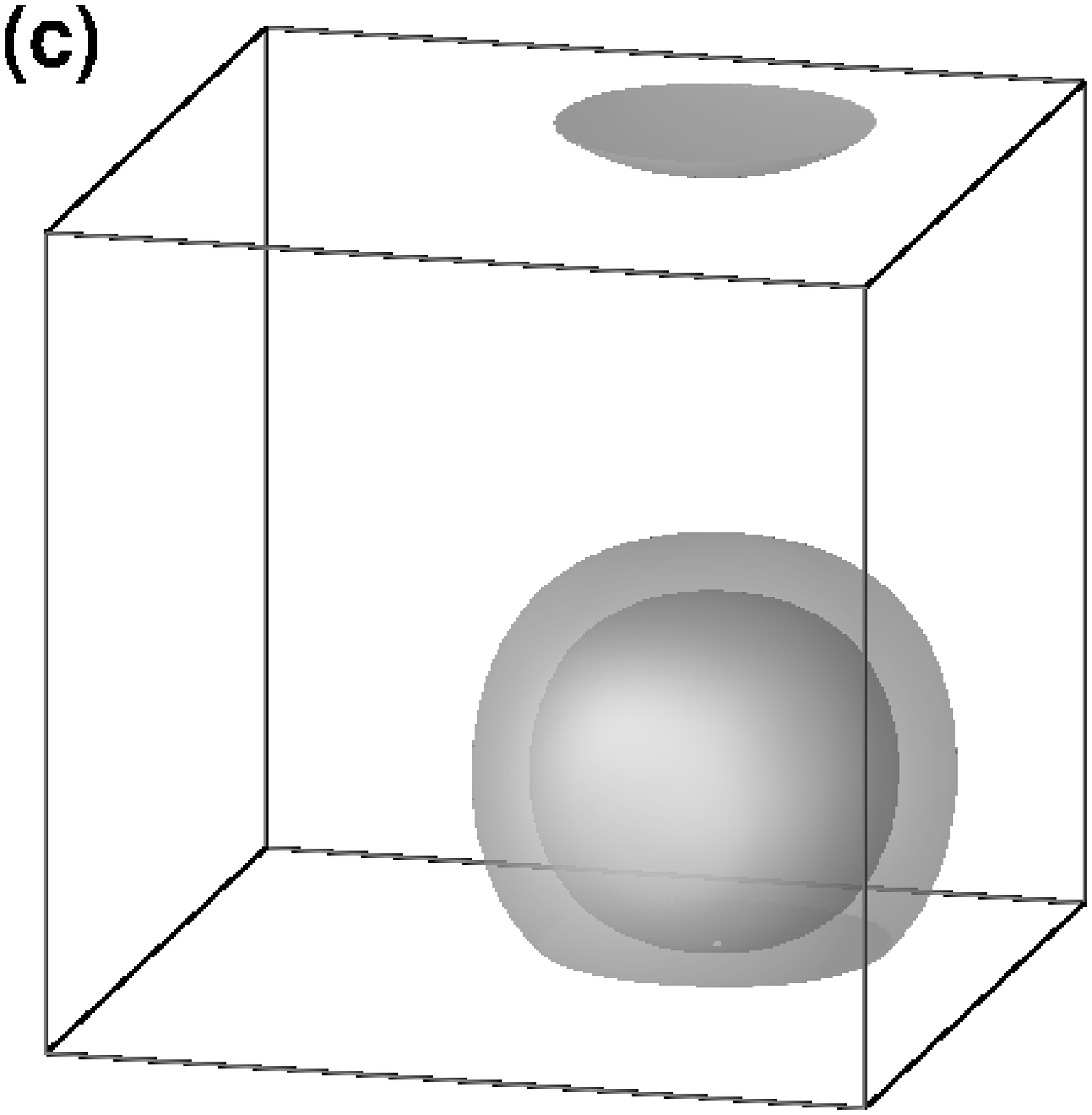}}
 \hspace{0.025\linewidth}
 {\includegraphics[width=0.45\linewidth,clip]{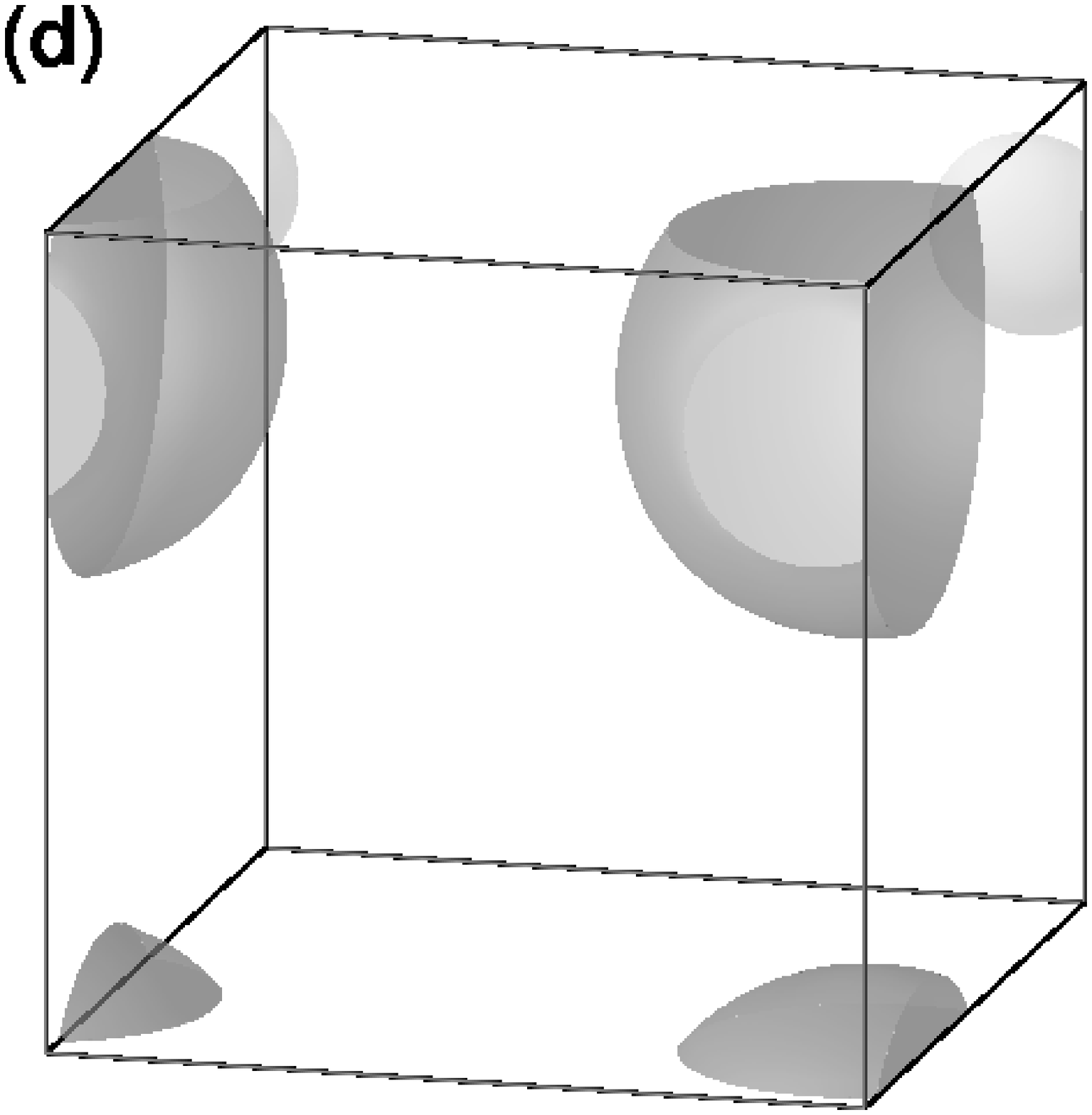}}
 \vspace{0.035\linewidth} {}
 {\includegraphics[width=0.9\linewidth,clip]{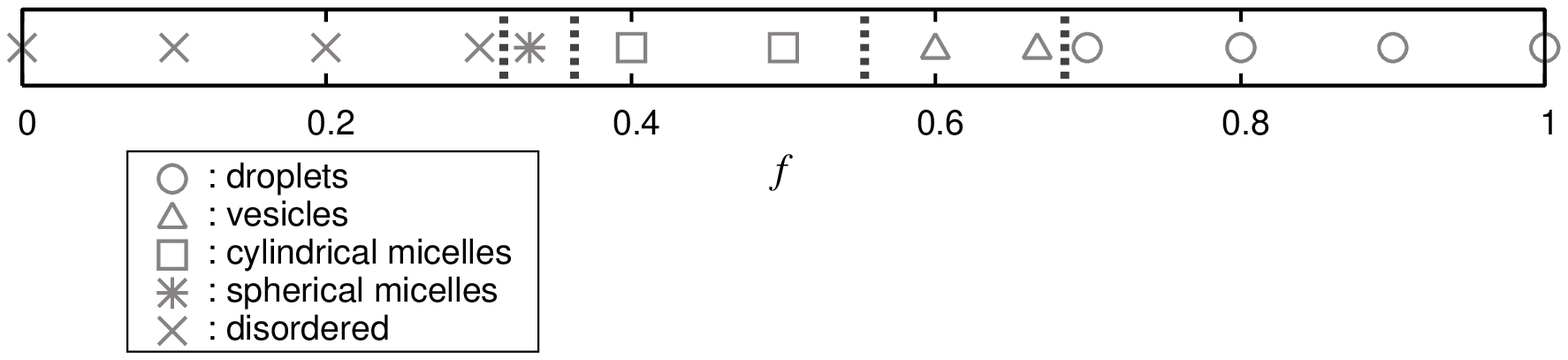}
 \caption{Phase diagram of AB diblock copolymer solutions.
 (a) $f_{B} = 1 / 3$, (b) $f_{B} = 0.5$, (c) $f_{B} = 0.6$, (d) $f_{B} = 1$.
 $N_{AB} = 20, N_{C} = 1, \bar{\phi}_{AB} = 0.1, \bar{\phi}_{C} = 0.9, \chi_{AB} = 1, \chi_{BC} = 1.75, \chi_{CA} =
-0.175$, system size: $32b \times 32b \times 32b$, lattice points: $64
 \times 64 \times 64$.}
 \label{solution_blockratio}
}
\end{figure}

To investigate the effect of the architecture of block copolymers, we
fixed the parameter except for the block ratio and performed several
simulations with different block ratio. The parameters are set to
$N_{AB} = 20, N_{C} = 1, \bar{\phi}_{AB} = 0.1, \bar{\phi}_{C} = 0.9, s\chi_{AB} = 1, \chi_{BC} = 1.75, \chi_{CA} =
-0.175$, system size: $32b \times 32b \times 32b$, lattice points: $64
\times 64 \times 64$
and the simulations have been done for block ratio $f_{B} = 0, 0.1, 0.2,
0.3, 1/3, 0.4,
0.5, 0.6, 2/3, 0.7, 0.8, 0.9, 1$. The result is shown in Figure \ref{solution_blockratio}.
It is seen that if the block ratio of the solvent-phobic subchain is too
short( i.e., if $f_B<1/3$), the block copolymer acts as the solvent-philic homopolymer. Thus the system is
homogeneous and the micellar structure is not formed. As increasing the block ratio of the solvent-phobic
subchian, spherical micelles and the cylindrical micelles are formed
(the relation between the morphology of micellar structure and the block
ratio is studied by Ohta and Nonomura for the strong segregation limit
\cite{Ohta-Nonomura-1998}).
As increasing the block ratio further, the vesicles are formed. In this
case the block ratio of solvent-phobic subchain is large, i.e. the block
copolymer is so-called crew-cut block copolymer. If one increase the
block ratio further, the block copolymer acts as the
solvent-phobic homopolymer and causes macro phase separation to form
droplets.

\subsection{AB Diblock Copolymer / A Homopolymer Blends}

We now study the AB diblock copolymer / A homopolymer blends. 
Like the case of solution, the AB diblock copolymer form micelles in the
matrix of A homopolymers.

From the technical view point, simulation of 
polymer blends is easier than the simulation of
polymer solutions since strong segregation is easily achieved
in polymer blends for small $\chi$ parameters: 
large $\chi$ parameter makes the interface sharper and causes
numerical stability problems at the interface.

\begin{figure}[p!]
 \centering
 {\includegraphics[width=0.9\linewidth,clip]{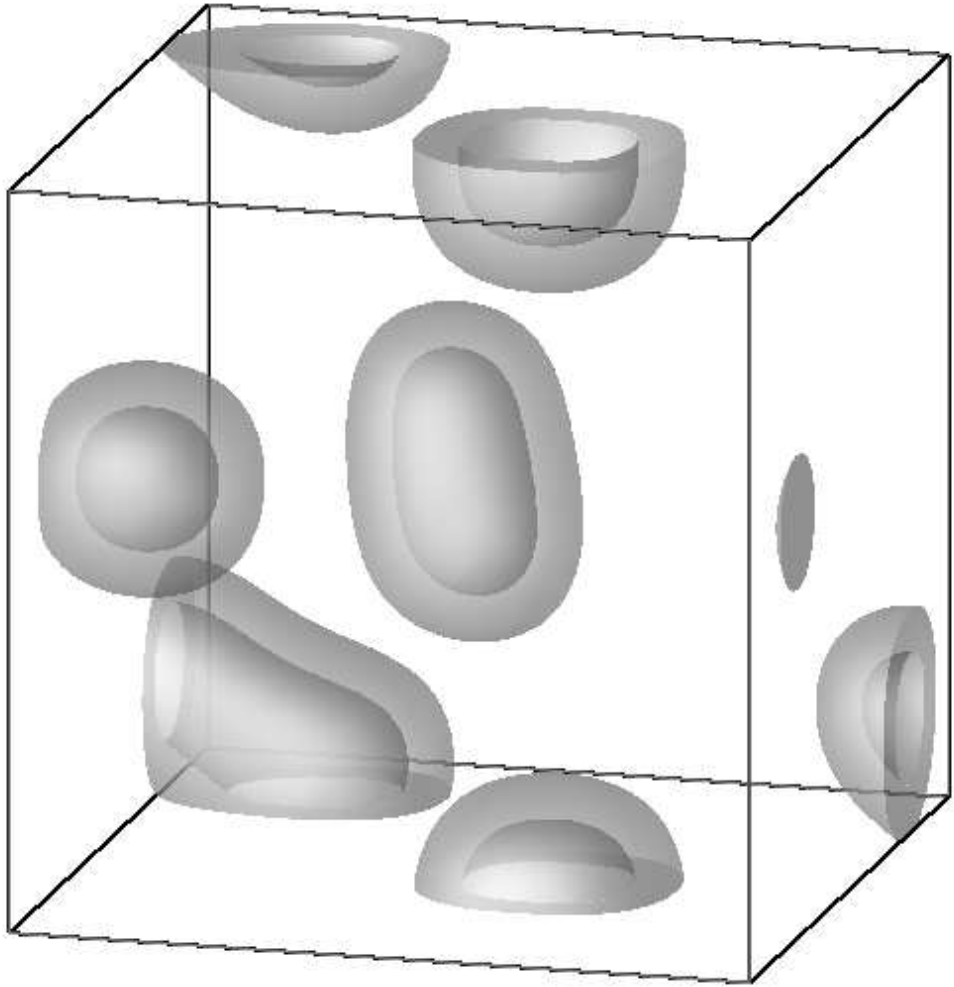}}
 \caption{A result of the simulation for the AB diblock copolymer / A
 homopolymer blend. The system size is $64 b \times 64 b \times 64 b$
 which involves  $128 \times 128 \times 128$ lattice points.  
 The parameters of polymers are $N_{AB} = 20, N_{A} = 10, 
 f_{A} = 1/3, f_{B} = 2/3, \chi_{AB} = 1$. 
 The volume fractions are
 $\bar{\phi}_{AB} = 0.1, \bar{\phi}_{A} = 0.9$. 
 The gray surfaces are isodensity surfaces for 
 $\phi_{B}(\bm{r}) = 0.5$.}
 \label{vesicle_simulation_large}
\end{figure}

Figure \ref{vesicle_simulation_large} shows an example of the simulation
result for a large system. Here the system involves $128 \times 128
\times 128$ lattice points (the system size is $64 b\times 64 b \times 64
b$), and the parameters are set to be $N_{AB} = 20, N_{A} = 10,
 f_{A} = 1/3, f_{B} = 2/3, \chi_{AB} = 1,
 \bar{\phi}_{AB} = 0.1, \bar{\phi}_{A} = 0.9$. In this case, vesicles
 can be observed more clearly than Figure \ref{solution}(c). This is
 considered to be due to the change of the strength of segregation (the
 segregation of this system is stronger than the system of Figure \ref{solution}(c)).

Various self organized structures can be formed depending on the
parameters such as the block ratio $f_{A}, f_{B}$, the volume fraction
$\bar{\phi_{AB}}$ and the $\chi$ parameter $\chi_{AB}$. To study the relation
between these parameters and the micellar structure, we conducted a
simulation of smaller system which involves $64 \times 64 \times 64$
lattice points. The polymerization index for the AB diblock copolymer
and the A homopolymer is set to $N_{AB} = 20, N_{A} = 10$, respectively.

\begin{figure}[p!]
 \centering
 {}
 {\includegraphics[width=0.45\linewidth,clip]{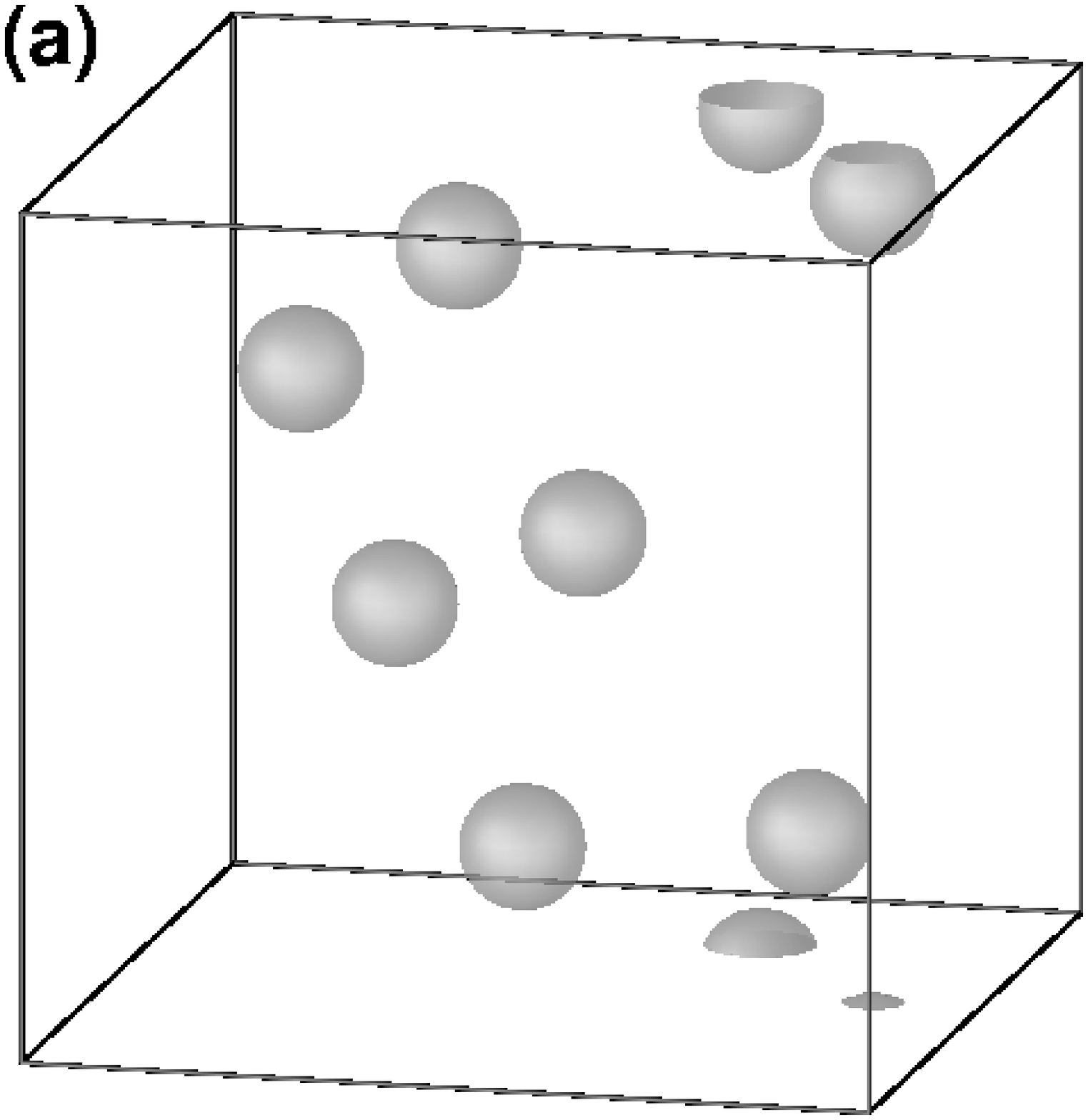}}
 \hspace{0.025\linewidth}
 {\includegraphics[width=0.45\linewidth,clip]{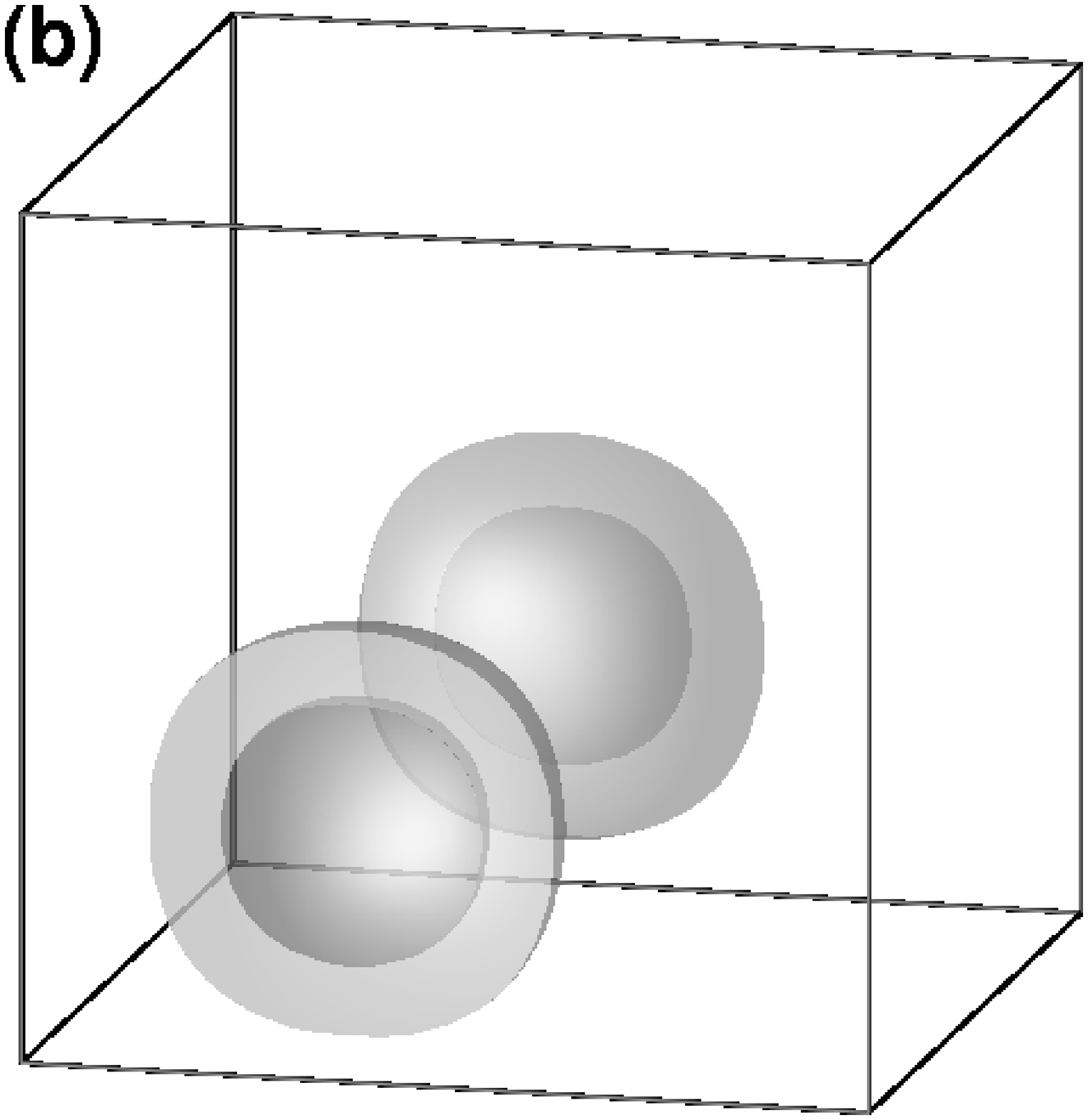}}
 \vspace{0.035\linewidth} {}
 {\includegraphics[width=0.45\linewidth,clip]{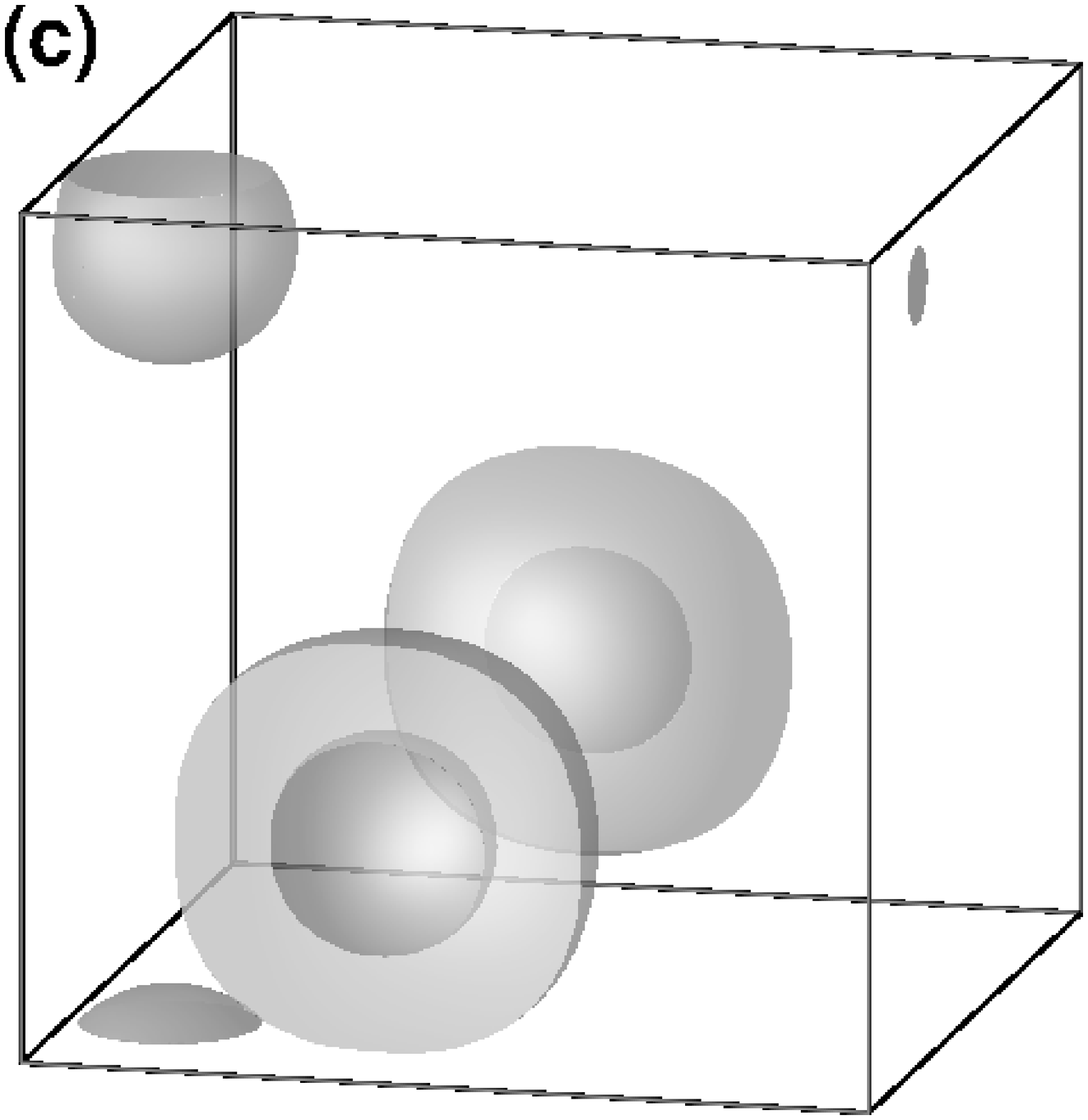}}
 \hspace{0.025\linewidth}
 {\includegraphics[width=0.45\linewidth,clip]{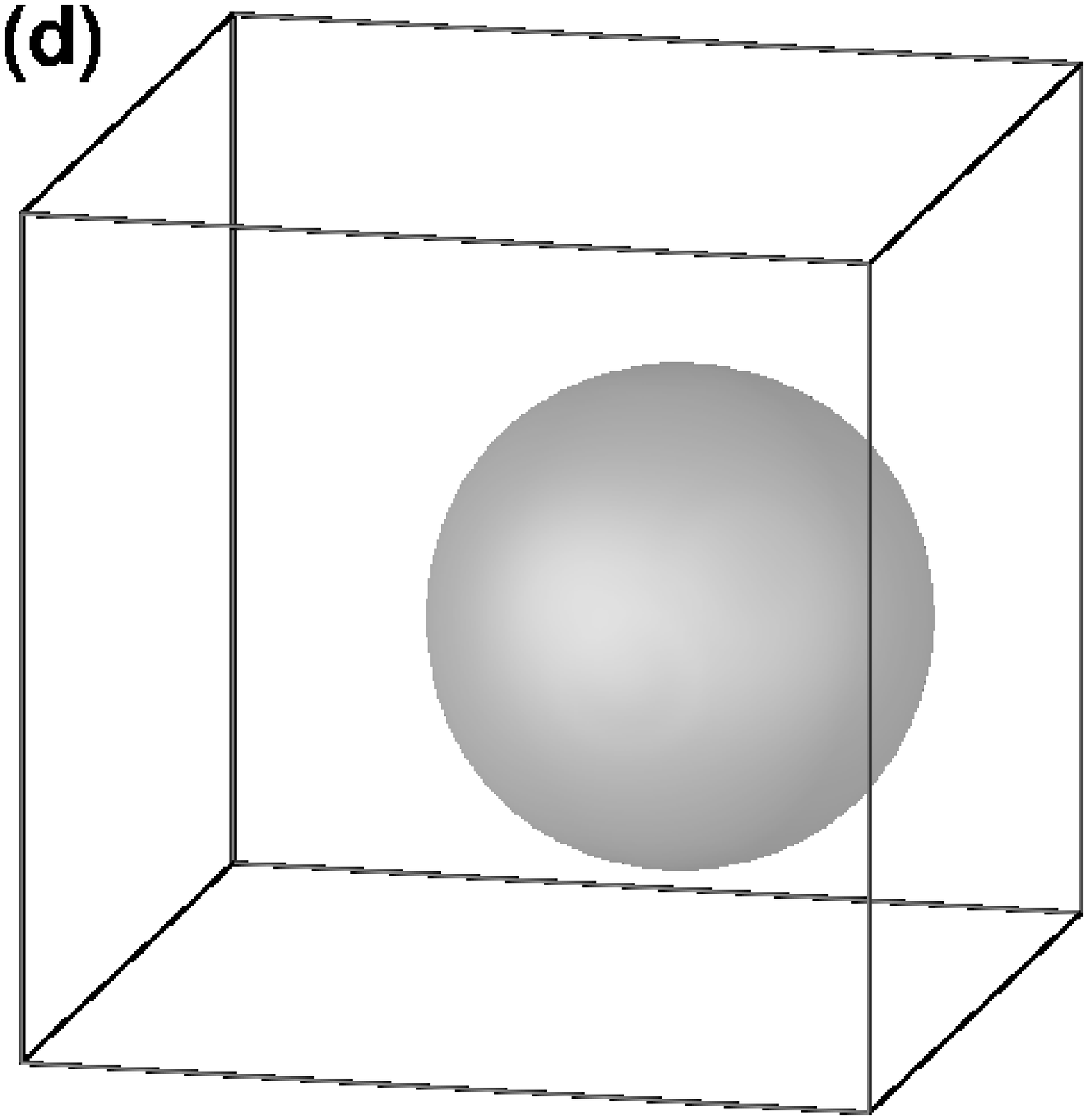}}
 \caption{Results of simulations with different block ratios. The
 system size is $32 b \times 32 b \times 32 b$. $N_{AB} = 20, N_{A} = 10, \bar{\phi}_{AB} = 0.1, \bar{\phi}_{A} = 0.9, \chi_{AB} = 1$.
 (a) $f_{B} = 1/3$,
 (b) $f_{B} = 0.5$, (c) $f_{B} = 2/3$, (d) $f_{B} = 1$.
 The gray surfaces are isodensity surfaces for $\phi_{B}(\bm{r}) = 0.5$.}
 \label{block_ratio_change}
\end{figure}

\begin{figure}[p!]
 \centering
 {}
 {\includegraphics[width=0.45\linewidth,clip]{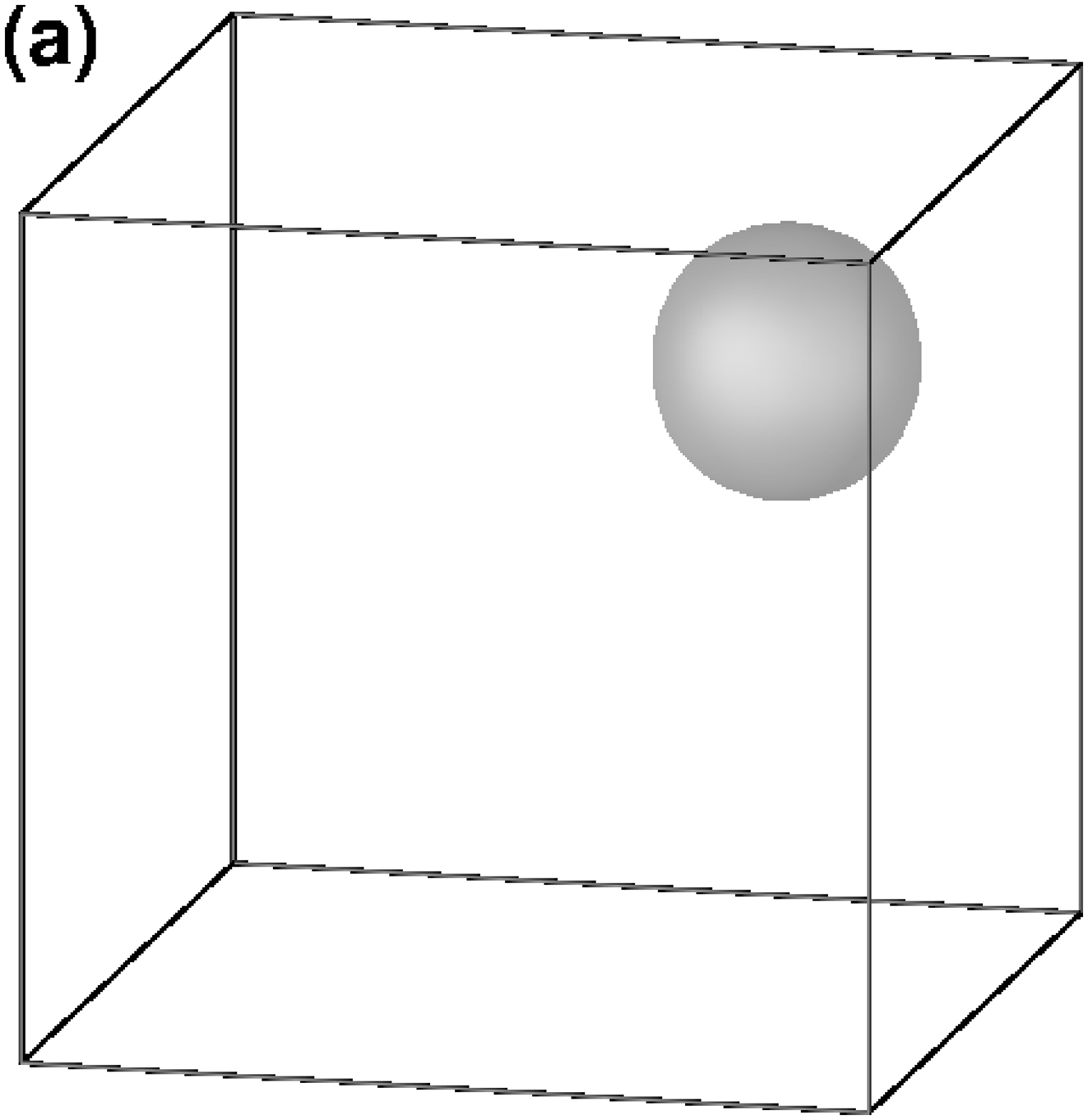}}
 \hspace{0.025\linewidth}
 {\includegraphics[width=0.45\linewidth,clip]{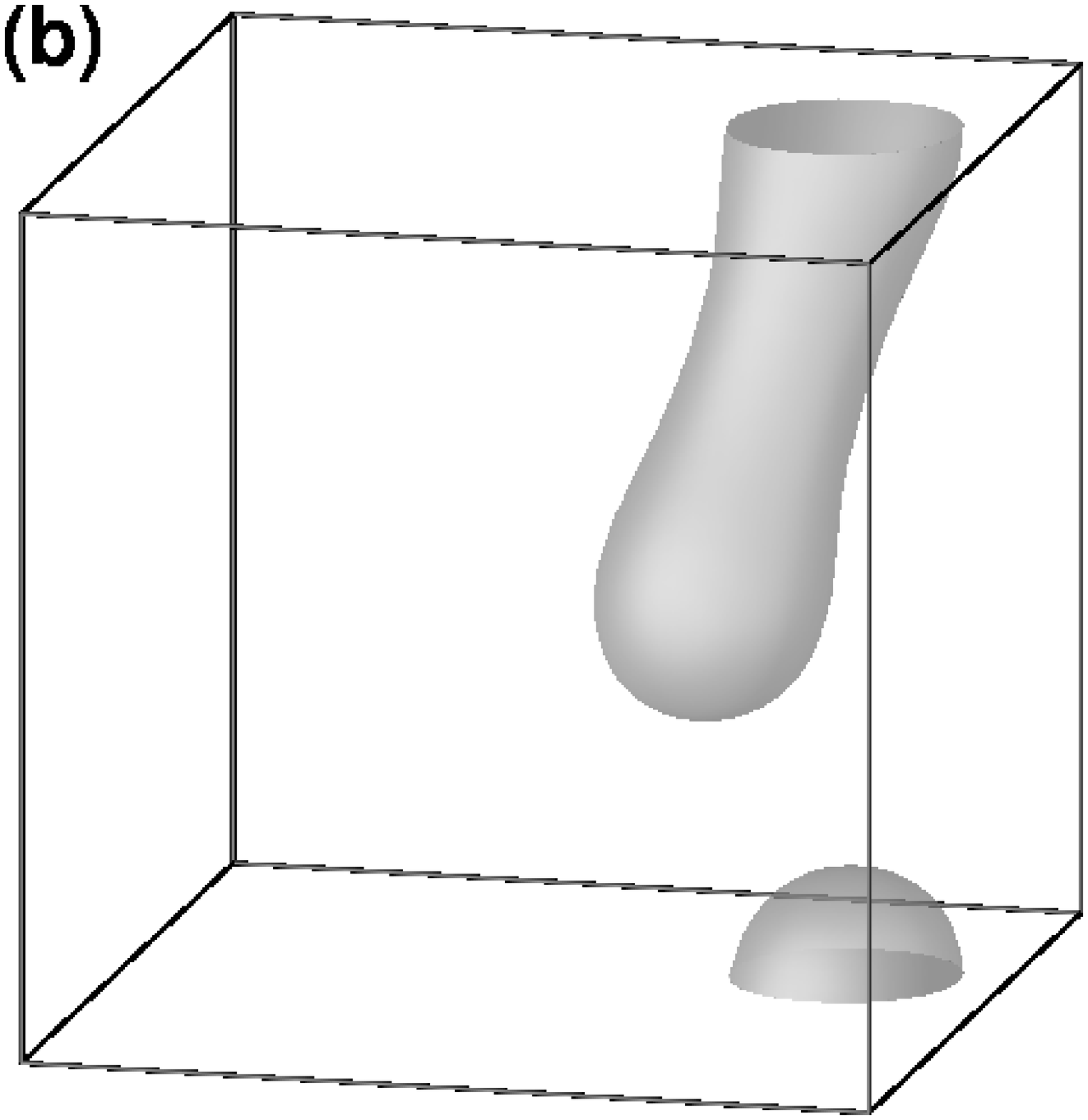}}
 \vspace{0.035\linewidth} {}
 {\includegraphics[width=0.45\linewidth,clip]{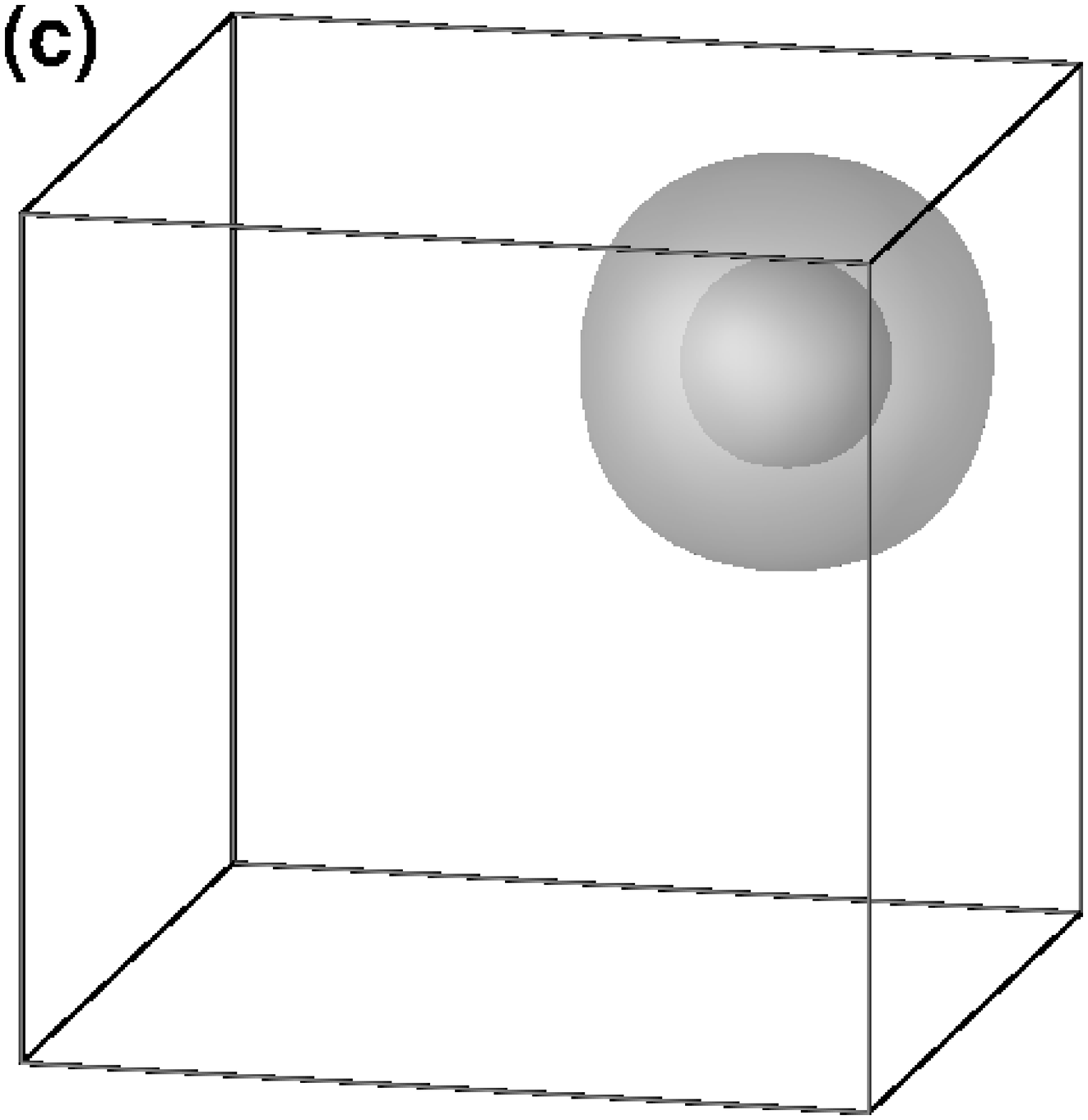}}
 \hspace{0.025\linewidth}
 {\includegraphics[width=0.45\linewidth,clip]{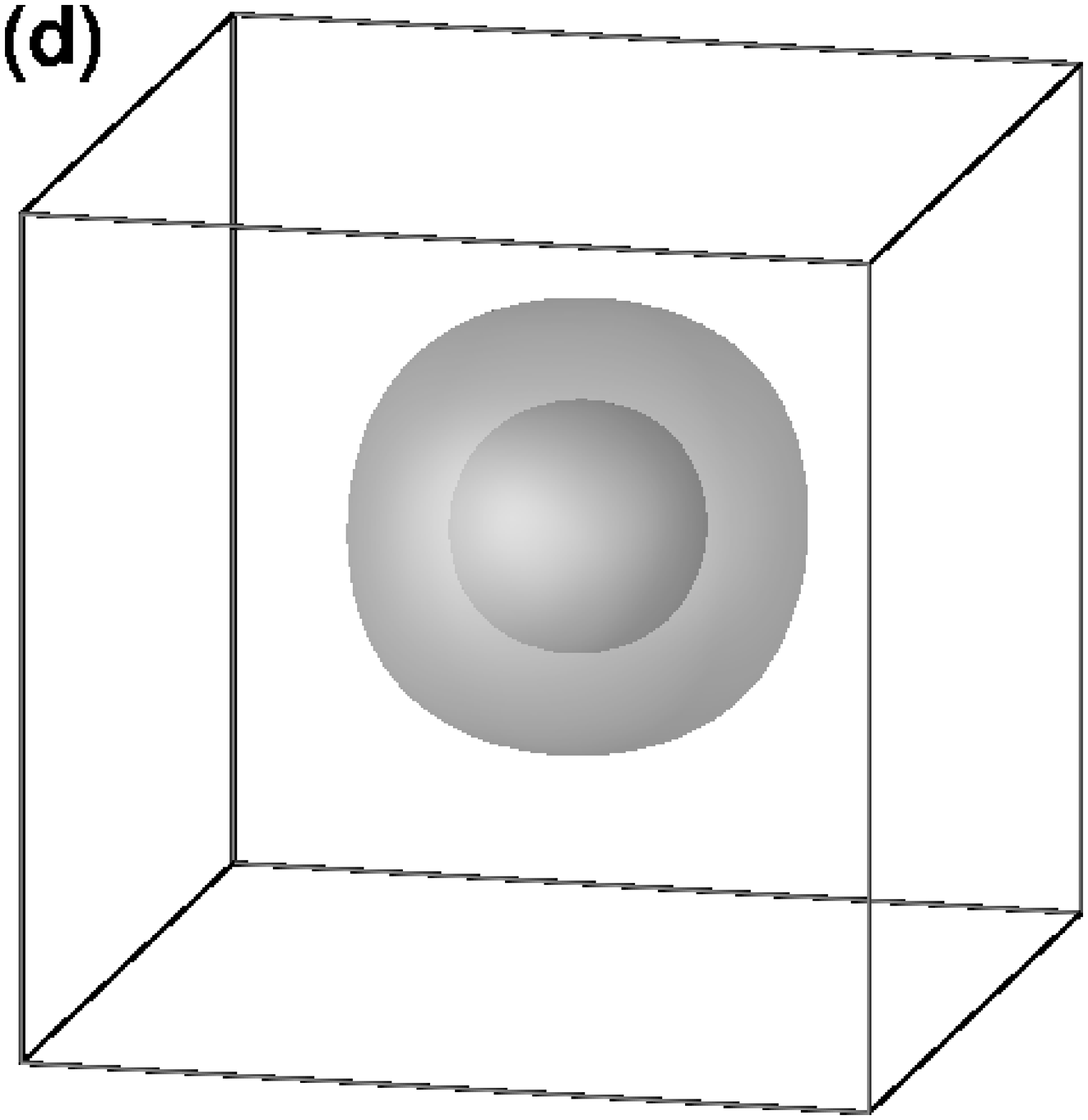}}
 \caption{Results of simulations with different volume fractions. The
 system size is $32 b \times 32 b \times 32 b$. $N_{AB} = 20, N_{A} = 10, f_{A} = 1/3, f_{B} = 2/3, \chi_{AB} = 1$. (a) $\bar{\phi}_{AB} = 0.025$,
 (b) $\bar{\phi}_{AB} = 0.0525$, (c) $\bar{\phi}_{AB} = 0.075$, (d)
 $\bar{\phi}_{AB} = 1$.
 The gray surfaces are isodensity surfaces for $\phi_{B}(\bm{r}) = 0.5$.}
 \label{volumefraction_change}
\end{figure}

\begin{figure}[p!]
 \centering
 {}
 {\includegraphics[width=0.45\linewidth,clip]{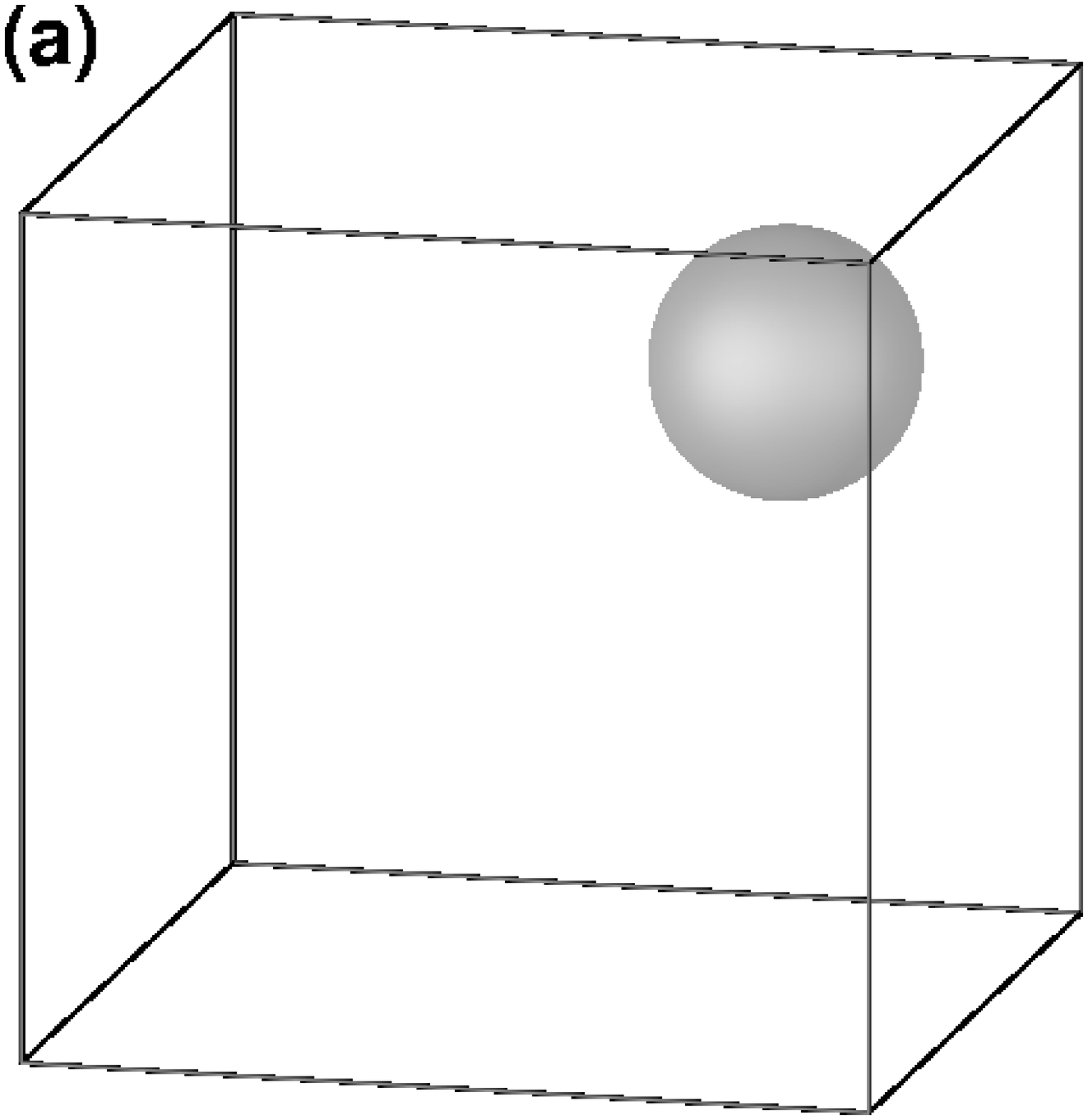}}
 \hspace{0.025\linewidth}
 {\includegraphics[width=0.45\linewidth,clip]{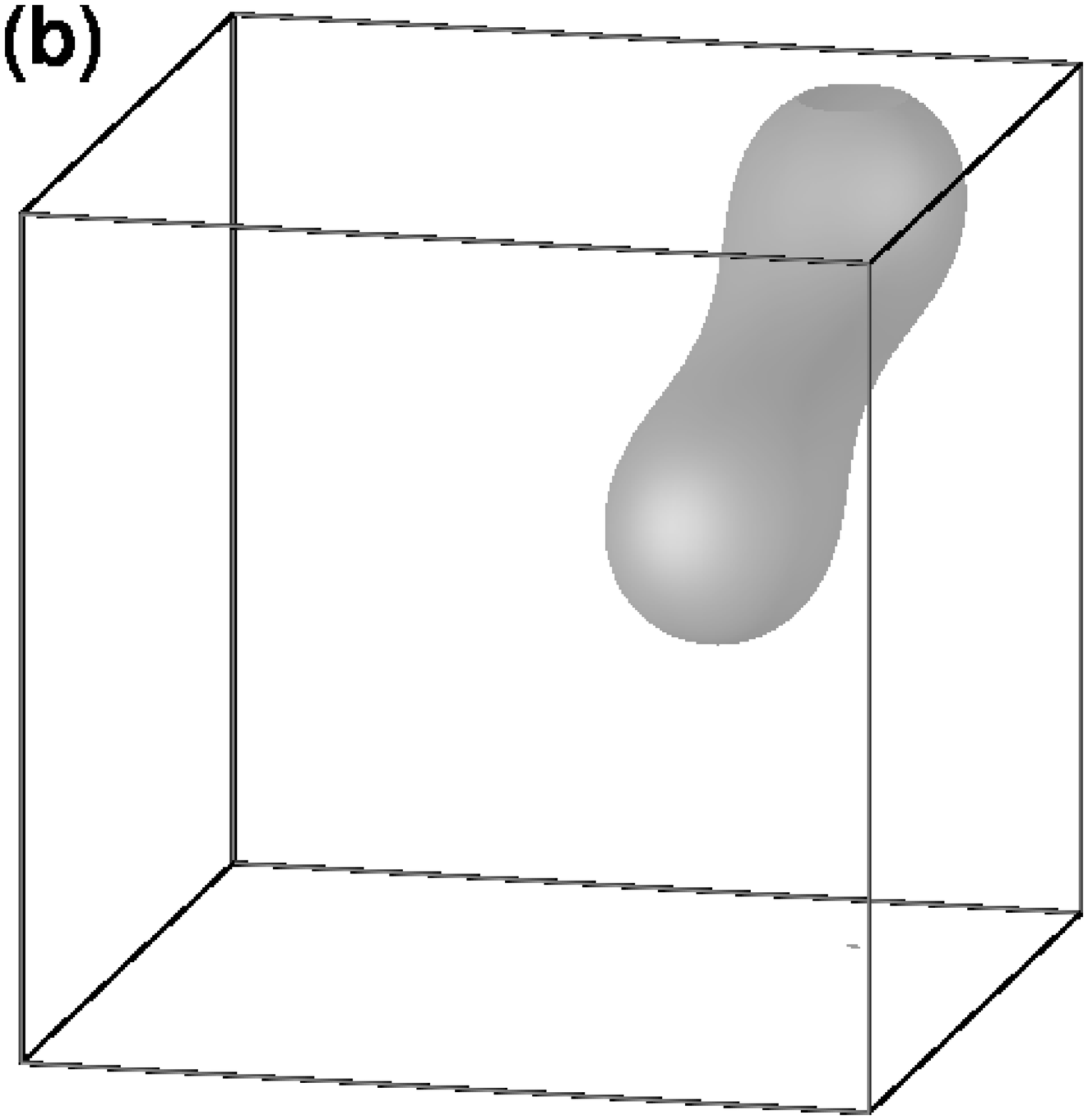}}
 \vspace{0.035\linewidth} {}
 {\includegraphics[width=0.45\linewidth,clip]{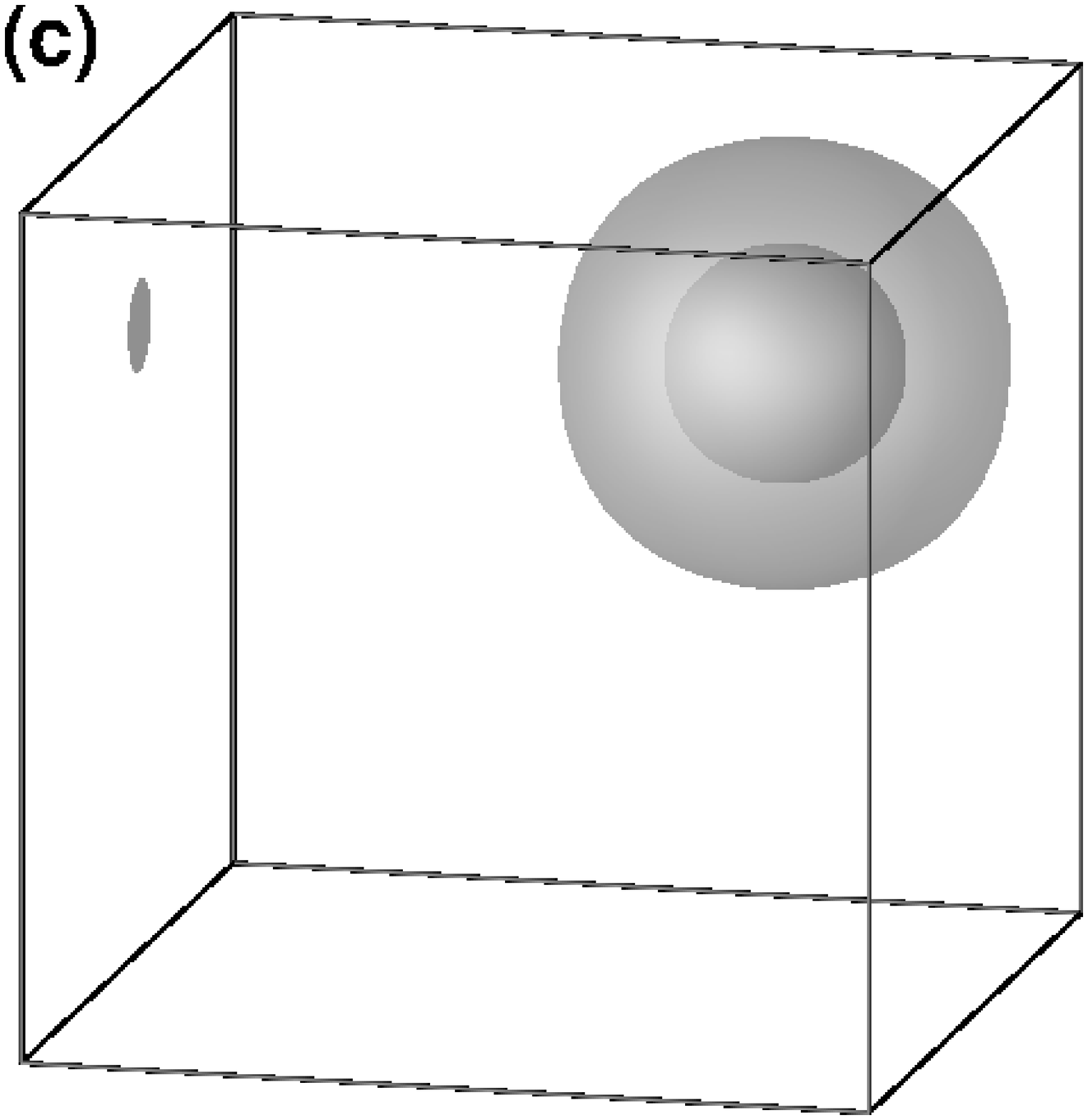}}
 \hspace{0.025\linewidth}
 {\includegraphics[width=0.45\linewidth,clip]{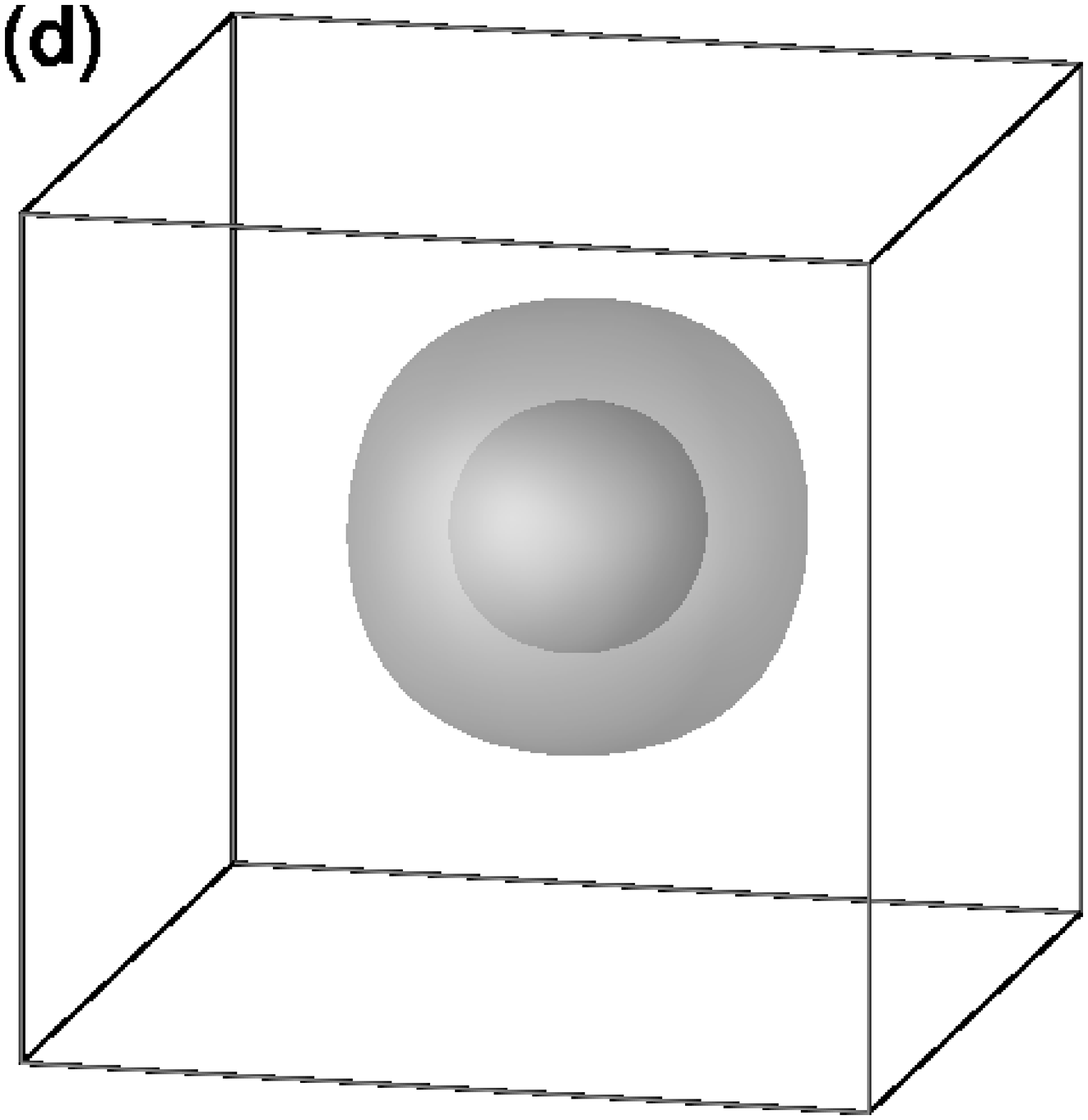}}
 \caption{Results of simulations with different $\chi$ parameters. The
 system size is $32 b \times 32 b \times 32 b$. $N_{AB} = 20, N_{A} = 10, f_{A} = 1/3, f_{B} = 2/3, \bar{\phi}_{AB} = 0.1, \bar{\phi}_{A} = 0.9$. (a) $\chi_{AB} = 0.0475$,
 (b) $\chi_{AB} = 0.525$, (c) $\chi_{AB} = 0.75$, (d) $\chi_{AB} = 1$.
 The gray surfaces are isodensity surfaces for $\phi_{B}(\bm{r}) = 0.5$.}
 \label{chi_parameter_change}
\end{figure}

Morphological change can be observed by chaining the parameters.
Figure \ref{block_ratio_change} shows the effect of the block ratio
$f_{A}, f_{B}$. Other parameters are set to $\bar{\phi}_{AB} = 0.1,
\bar{\phi}_{A} = 0.9,
\chi_{AB} = 1$.  If the B block is small (Figure
\ref{block_ratio_change}(a), $f_{B} = 1/3$), the diblock copolymer forms
a spherical micelle which consists of
inner core made of B block and the outer corona made of A block. As the
fraction of the B block increases, the spherical micelles become
unstable, and vesicles consisting of a bilayer of the block copolymers is
formed.  (Figure \ref{block_ratio_change}(b), $f_{B} = 0.5$). The thickness of the B
block in the bilayer increases with the increase of B block
ratio. Finally, when $f_{B}$ becomes equal to unity, the macro-phase
separation between A homopolymer and B homopolymer is
observed. (Fig. \ref{block_ratio_change}(d), $f_{B} = 1$).
This behavior is just the same as one of the block copolymer solutions.
Figure \ref{volumefraction_change} shows the effect of
$\bar{\phi}_{AB}$, the volume fraction of diblock copolymers. Other
parameters are set to $f_{A} = 1/3, f_{B} = 2/3, \chi_{AB} = 1$.  As the volume
fraction of the diblock copolymer increases, the micellar structures
change from the spherical micelles to the cylindrical micelles and
vesicles.
Figure \ref{chi_parameter_change} shows the effect of the $\chi$
parameter between the monomer A and monomer B. Other parameters are set
to $f_{A} = 1/3, f_{B} = 2/3$, $\bar{\phi}_{AB} = 0.1, \bar{\phi}_{A} = 0.9$. As the $\chi$ parameter
increases, (i.e., as the antagonicity between the monomers A and B
increases), the structures change from the spherical micelles to the
cylindrical micelles and then to vesicles. This behavior is just like
the case of the volume fraction change.

These results are considered to be consistent with the experimental results 
of Eisenberg et al..
\cite{Shen-Eisenberg-1999,Shen-Eisenberg-2000,Choucair-Eisenberg-2003}.
They studied the morphology of diblock copolymers in mixed solvent
(water plus dioxane) for various block copolymer concentration and for
various solvent composition, and observed the change of the micellar
shape from sphere to cylinder and then to vesicles as the water content
of the solvent increases.  If one regard the change of the water content
as the change of the $\chi$ parameter between the solvent and the subchain of block
copolymer which forms core, our
results are qualitatively in agreement with their experiments.

\begin{figure}[p!]
 \centering
 {\includegraphics[width=0.9\linewidth,clip]{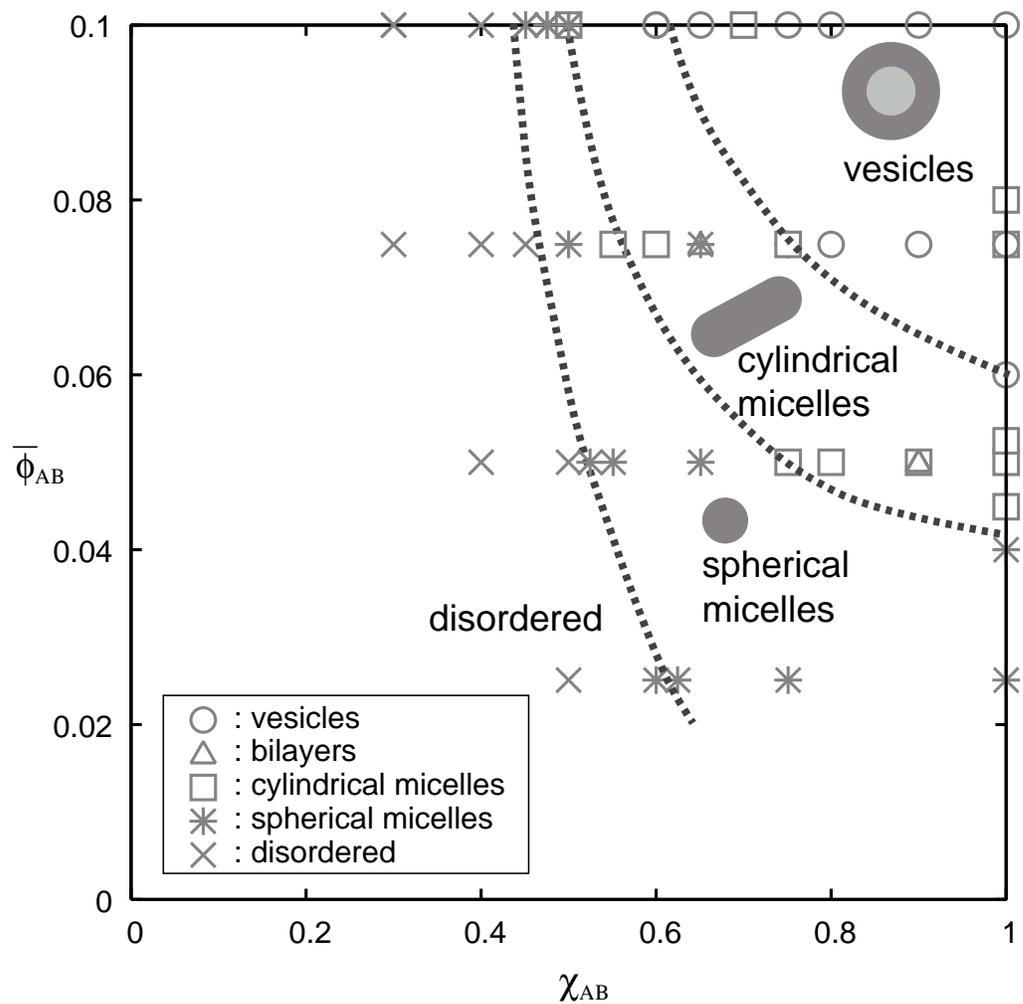}
 \caption{Phase diagram of AB diblock copolymer / A homopolymer blends.
 The plotted symbols (circles, squares, etc.) correspond to the result
 of simulations with various $\chi_{AB}$ and $\bar{\phi}_{AB}$.
 The bilayer phase means open, disk-like micelles (actually it is observed by
 experiment \cite{Shen-Eisenberg-2000}).}
 \label{phase_diagram}
}
\end{figure}

In order to demonstrate the similarity between the experimental
results and our simulations, we conducted the simulation in the parameter
space of $\bar{\phi}_{AB}$ and $\chi_{AB}$.  Fig. \ref{phase_diagram} shows the
result. 
The figure is in qualitatively agreement with Fig. 6 in
Ref.~\citen{Shen-Eisenberg-1999}. Note that our simulations was not done for
large systems and the size of the simulation cell may affect 
the final micellar structure (the finite size effect or 
the effect of the periodic boundary conditions).

%
%

\section{Discussion}
The micellar structures, including vesicles can be formed for the AB
diblock copolymer solutions by using the
free energy model eq~\eqref{freeenergy}.
The diblock copolymer density is dilute in our systems and the diblock
copolymer strongly localized at the micellar structures (i.e. the system is
the strong segregation).
Note that such systems are difficult to treat by the
previous density functional approaches.
Our free energy model can handle
these systems qualitatively correct and the simulation with our model is much
faster than the real-space SCF simulation. 

The dynamic vesicle formation process, however, cannot be discussed by
our numerical scheme because our scheme does not satisfy the local
conservation of mass. This is in contrast to the particle methods (BD
\cite{Noguchi-Takasu-2001} and DPD \cite{Yamamoto-Maruyama-Hyodo-2002}).
Nevertheless it is instructive to study the process of the micellar
formation based on our model.  

\begin{figure}[p!]
 \centering
 {}
 {\includegraphics[width=0.275\linewidth,clip]{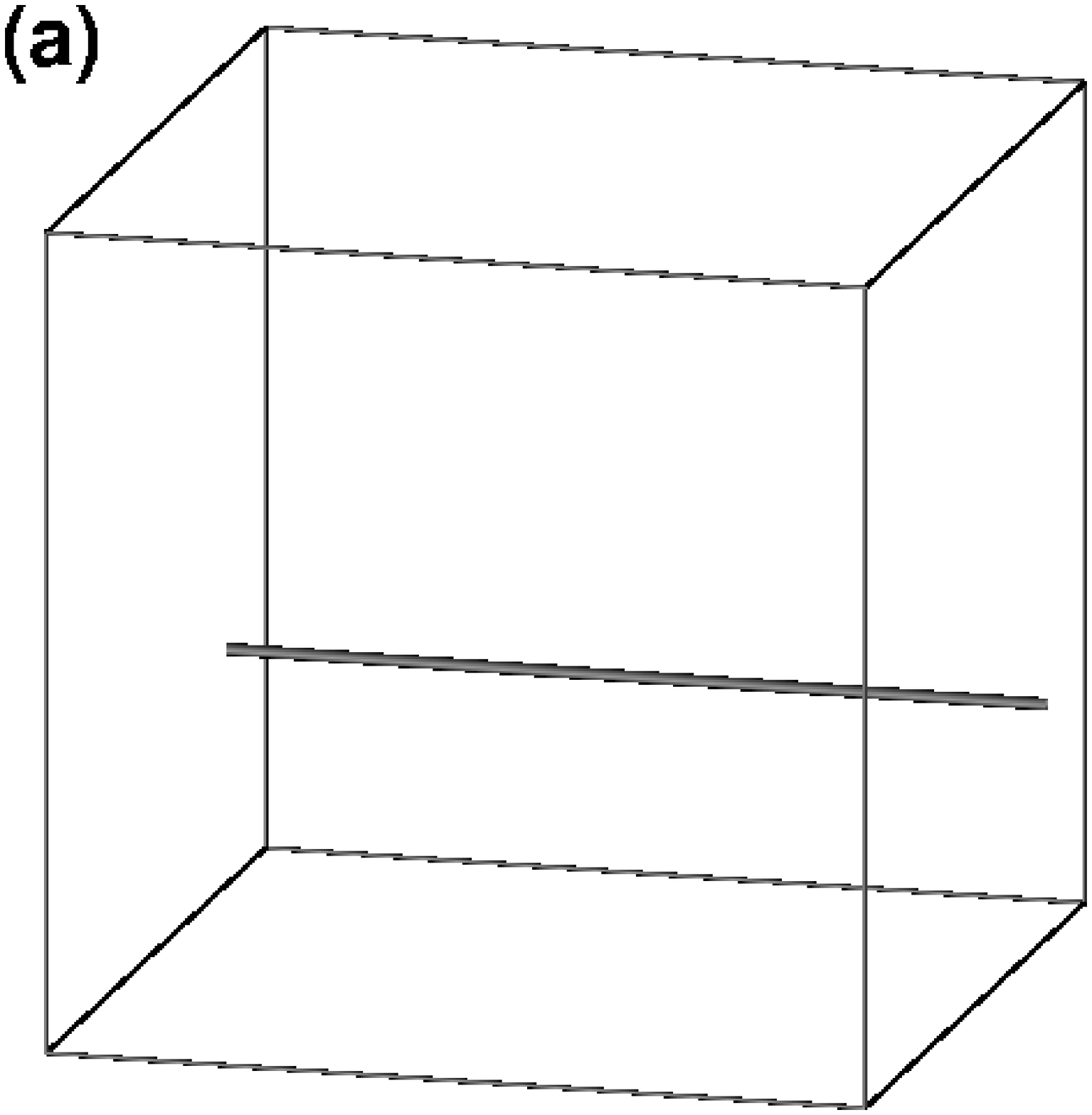}}
 \hspace{0.025\linewidth}
 {\includegraphics[width=0.4\linewidth,clip]{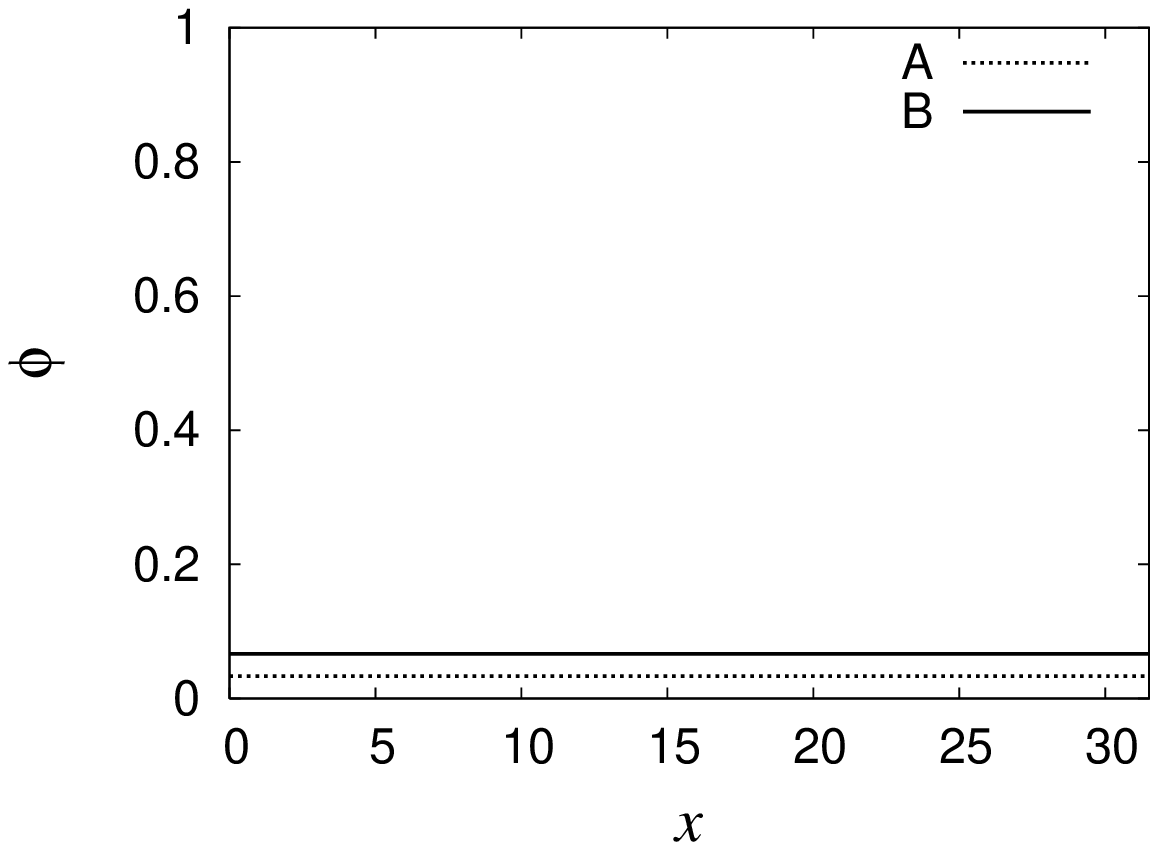}} {}
 \hspace{0.1\linewidth}
 {\includegraphics[width=0.275\linewidth,clip]{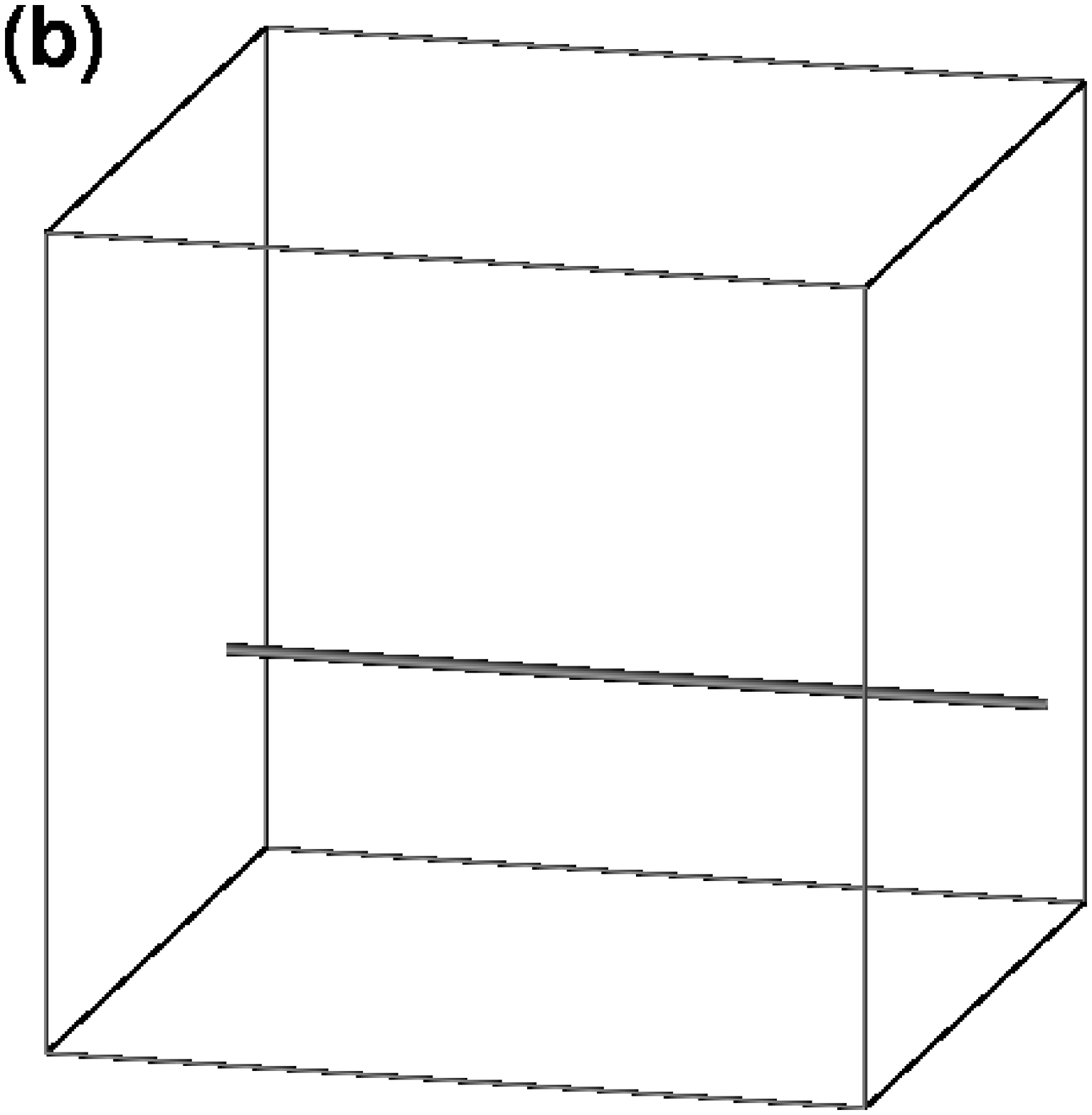}}
 \hspace{0.025\linewidth}
 {\includegraphics[width=0.4\linewidth,clip]{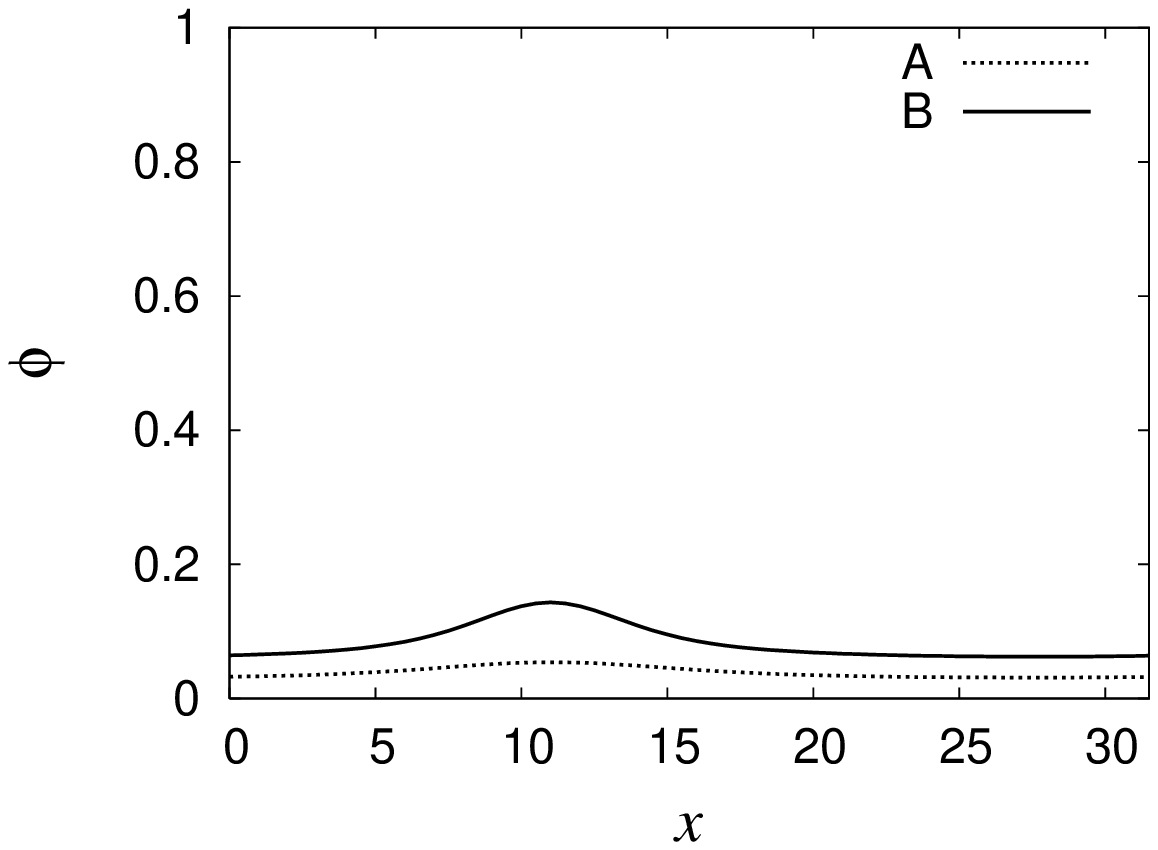}} {}
 \hspace{0.1\linewidth}
 {\includegraphics[width=0.275\linewidth,clip]{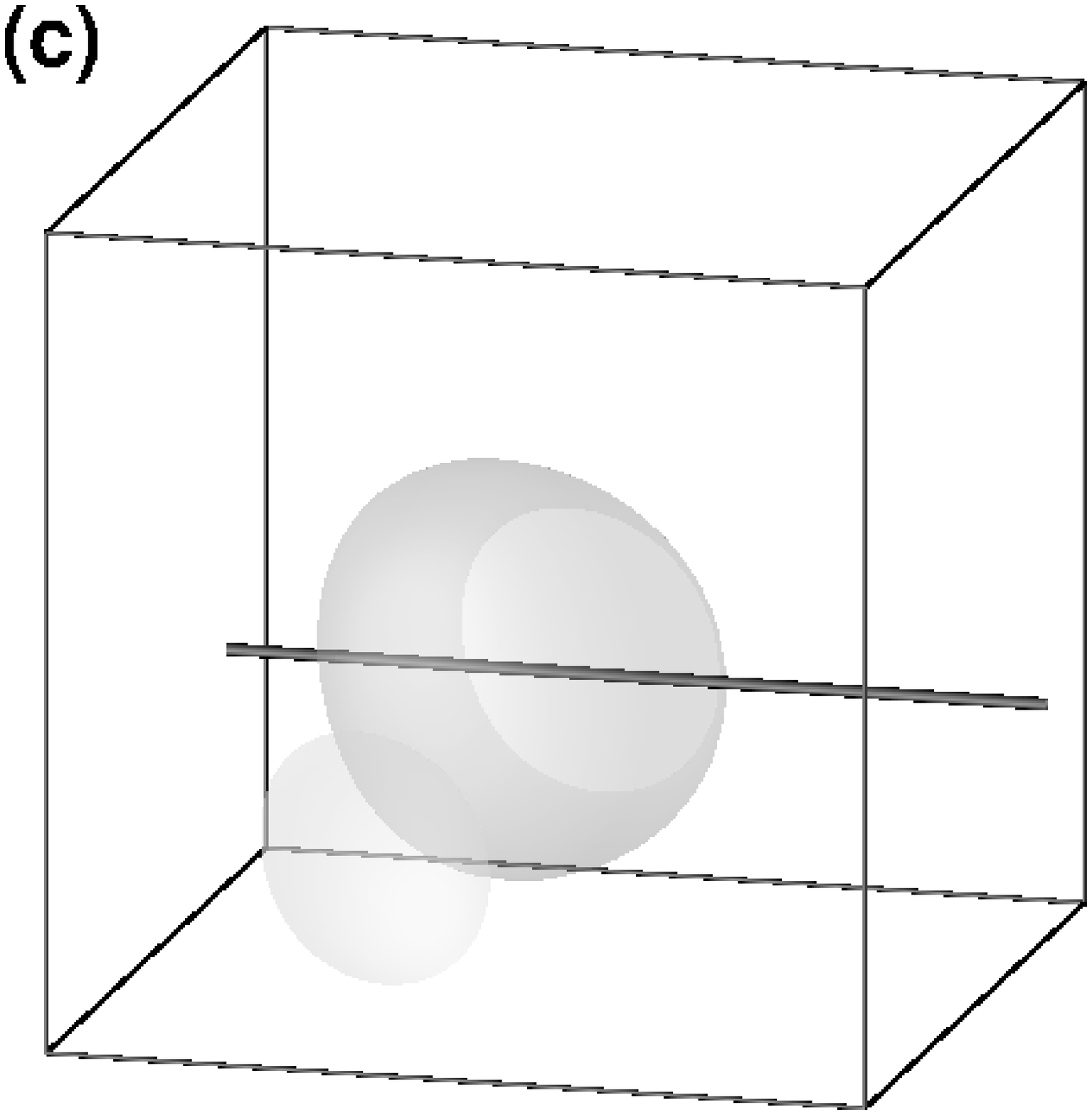}}
 \hspace{0.025\linewidth}
 {\includegraphics[width=0.4\linewidth,clip]{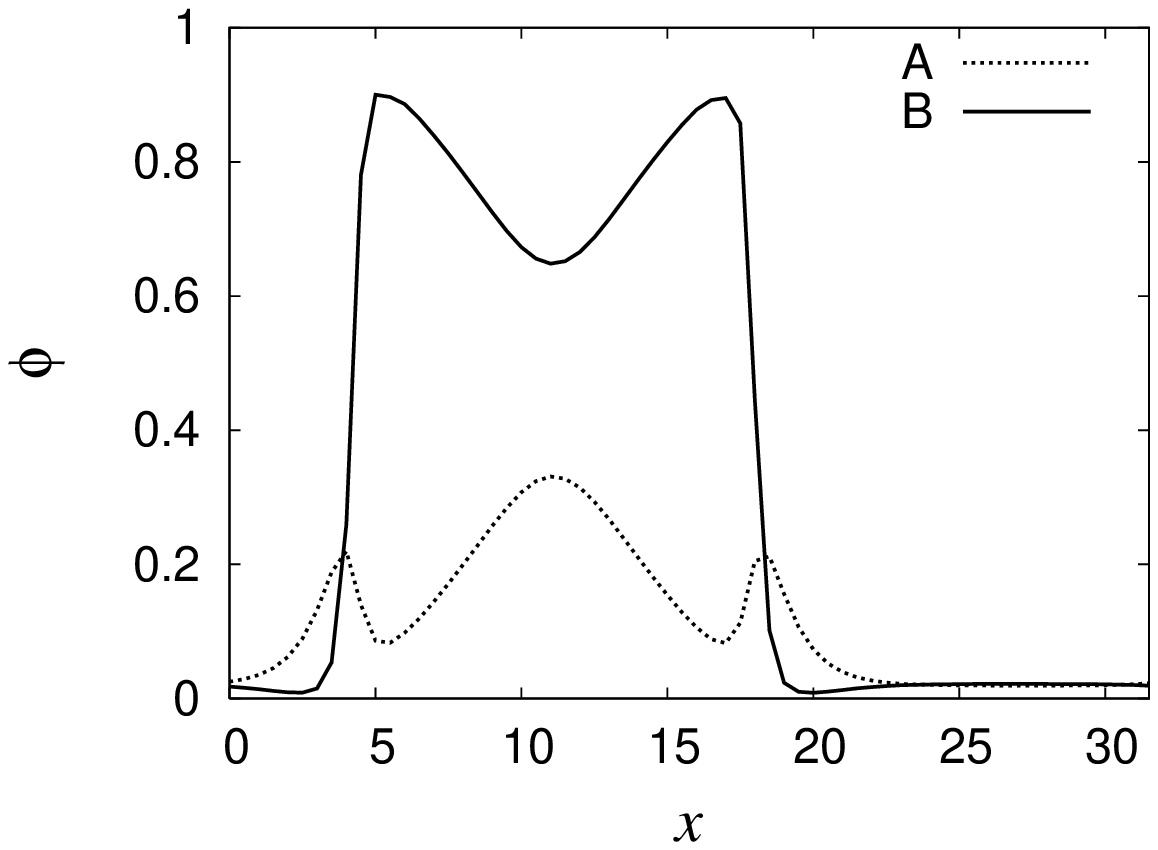}} {}
 \hspace{0.1\linewidth}
 {\includegraphics[width=0.275\linewidth,clip]{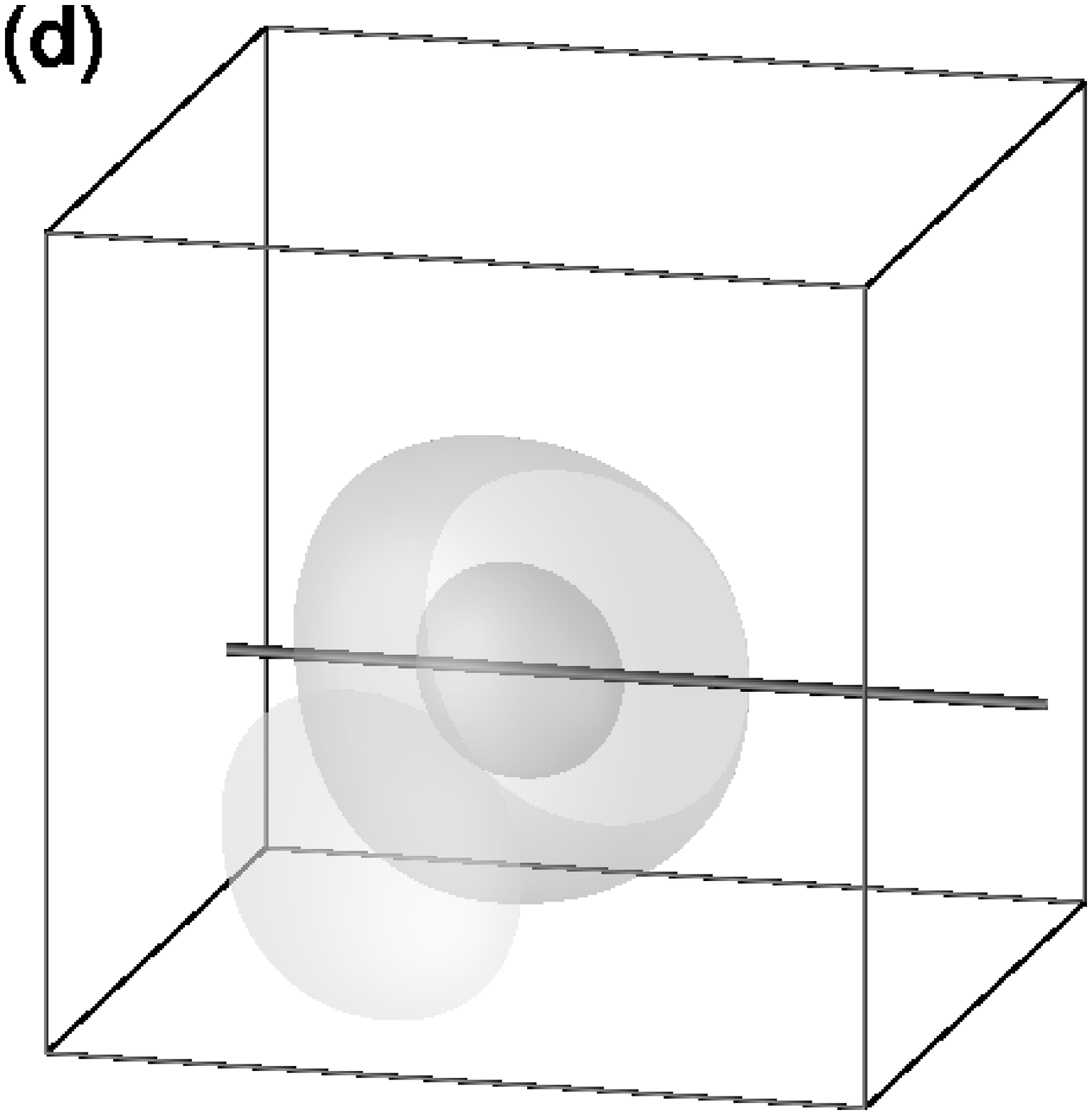}}
 \hspace{0.025\linewidth}
 {\includegraphics[width=0.4\linewidth,clip]{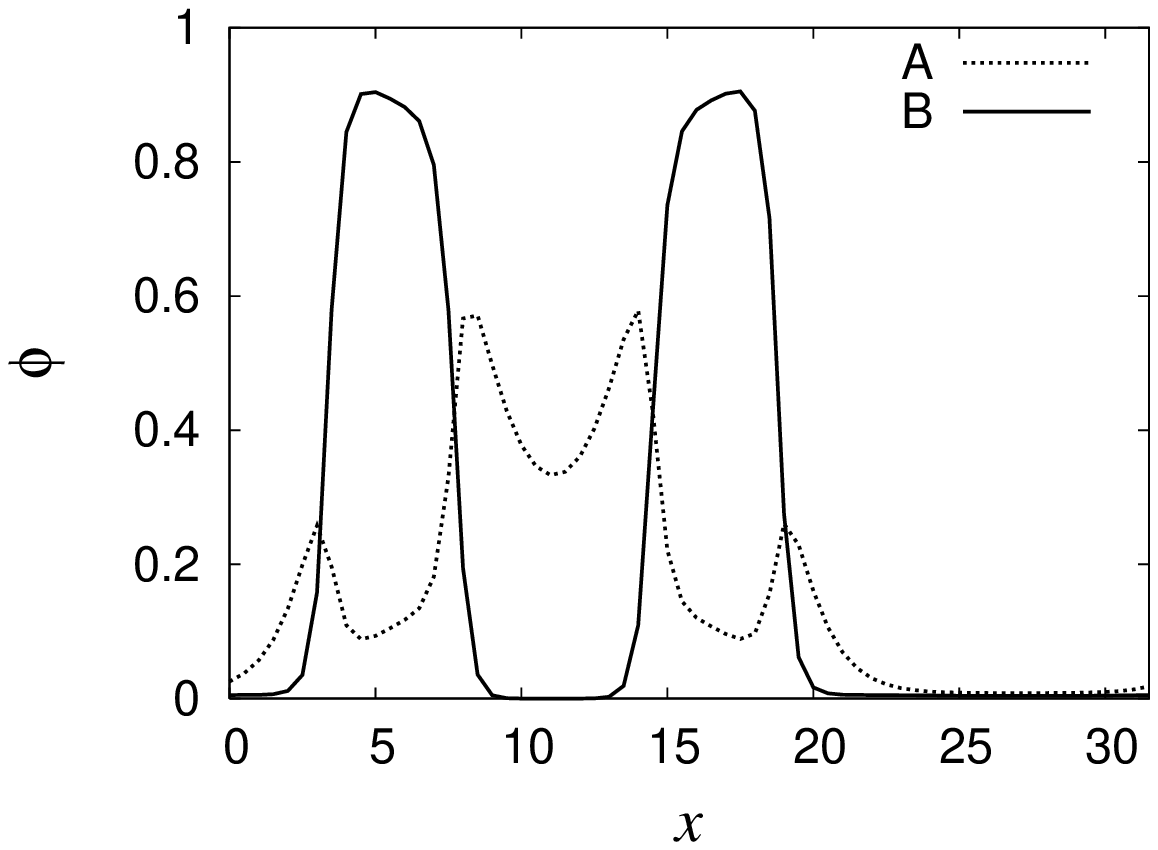}} {}
 \hspace{0.1\linewidth}
 {\includegraphics[width=0.275\linewidth,clip]{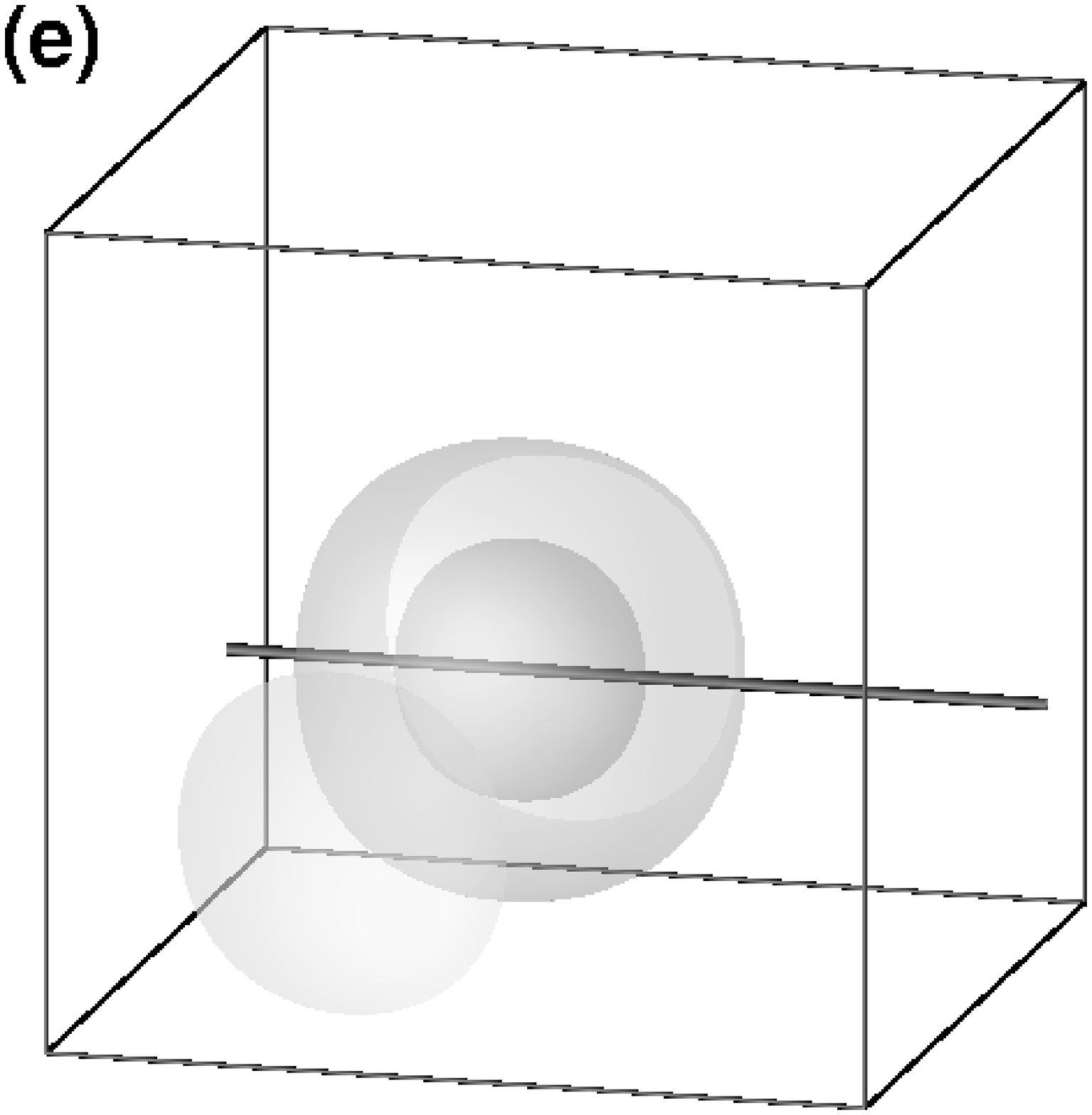}}
 \hspace{0.025\linewidth}
 {\includegraphics[width=0.4\linewidth,clip]{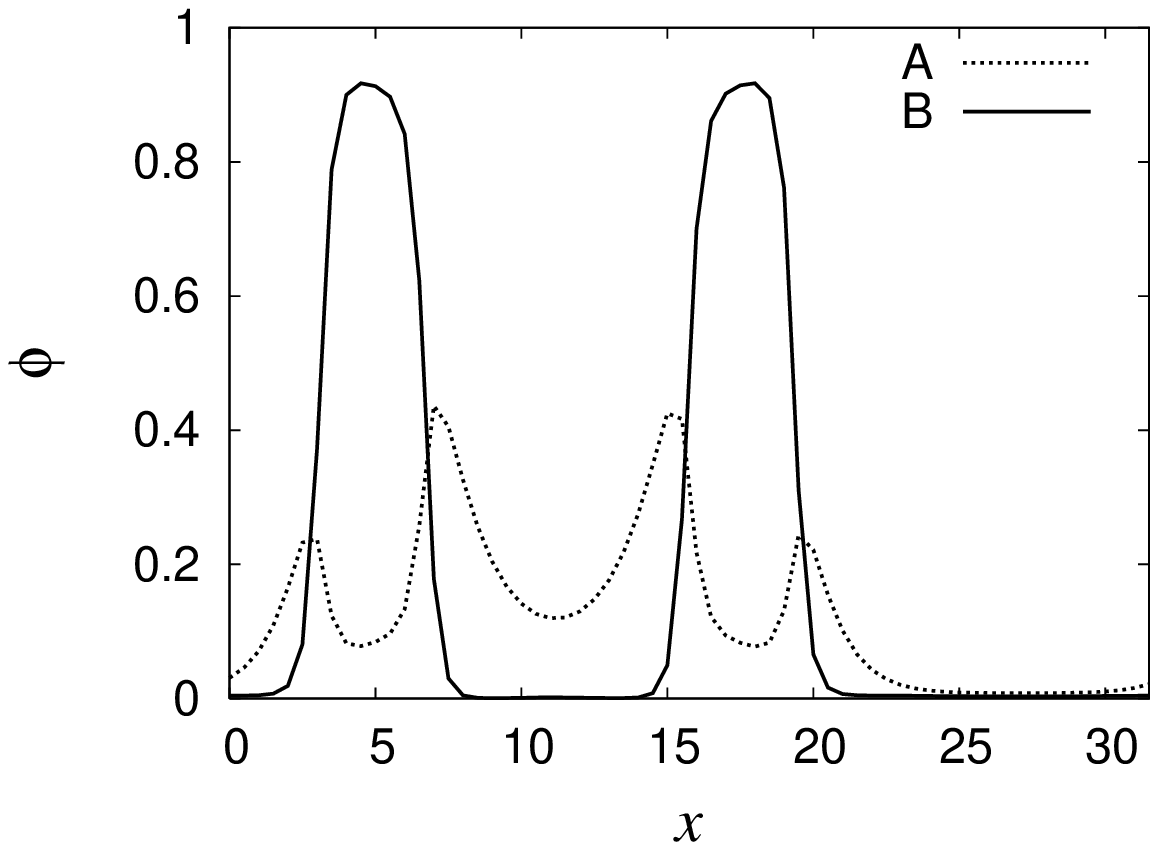}} {}

 \caption{Snapshot of vesicle formation process for the AB diblock
 copolymer / A homopolymer blend. Number of iteration
step = $0, 470, 500, 530, 990$ for (a), (b), (c), (d), (e),
 respectively. $N_{AB} = 20, N_{A} = 10,
 f_{A} = 1/3, f_{B} = 2/3, \chi_{AB} = 1,
 \bar{\phi}_{AB} = 0.1, \bar{\phi}_{A} = 0.9$, system size: $32b \times
 32b \times 32b$, lattice points: $64 \times 64 \times 64$ (The
 parameters are the same as Figure \ref{vesicle_simulation_large} except
 for the system size.)}
 \label{vesicle_formation}
\end{figure}

The vesicle formation process for the AB diblock copolymer / A homopolymer
blend by our scheme is shown in Figure
\ref{vesicle_formation} ($N_{AB} = 20, N_{A} = 10, 
 f_{A} = 1/3, f_{B} = 2/3, \chi_{AB} = 1,
 \bar{\phi}_{AB} = 0.1, \bar{\phi}_{A} = 0.9$, system size: $32b \times
 32b \times 32b$, lattice points: $64 \times 64 \times 64$,
 $\omega^{(1)} = 0.5$). It is observed that the nucleus of micellar
structure is first formed by the association of block copolymers
(the matrix polymer is expelled from the micellar structure
at this stage). The micellar structure then 
grows taking solvent into its core from the surrounding, and
finally the vesicle is formed.  

This process is the same as that observed by He et
al.\cite{He-Liang-Huang-Pan-2004} in their 2 dimensional simulation using the
SCF method. (They also used the non-conserving scheme for the
evolution of the local density.)
According to the scenario of He et al., the nucleus
of micellar structures grows first. If the nucleus grows sufficiently
large, the nucleus takes solvent into the core because the 
hydrophilic subchains localized at the large micelles are 
energetically unfavorable. Finally the micellar
structure changes their shape from the sphere-like structure into
shell-like structure and thus vesicles are formed.

Therefore our vesicle formation process is the same as one of He et al..
This agreement implies that the schemes which does not satisfy the
local conservation can form vesicles easily compared with the 
schemes which is based on the dynamics and
satisfy the local conservation (for example, the Cahn-Hilliard type TDGL
equation or the dynamic SCF method \cite{Fraaije-1993}).
The dynamic vesicle formation process based on the continuous field
model simulation is the future work.

%
%

\section{Conclusion}
We have shown that the self-organized structure of block
copolymers in solvent can be predicted by the density functional theory
using the free energy proposed in Ref.~\citen{Uneyama-Doi-2005}. This is
the first 3 dimensional simulation of vesicle formation by using the
continuous field model.
Since the theory can take into account of actual structure of polymers
(degree of polymerization, block ratio, topological structure), it will be
useful to understand and to predict the micellar structure of surfactant
systems.


\appendix
\section*{Appendix}
%
%

\section{Form of Matrices $A_{ij}, C_{ij}$}
\label{form_of_matrices_a_c}
The matrices $A_{ij}, C_{ij}$ for diblock copolymer is represented as
follows \cite{Uneyama-Doi-2005}.
\begin{align}
 A_{ij} & = \frac{9}{N_{AB}^{2} b^{2} f_{A}^{2} f_{B}^{2}}
 \begin{bmatrix}
  f_{B}^{2} & - f_{A} f_{B} \\
  - f_{A} f_{B} & f_{A}^{2}
 \end{bmatrix} \\
 C_{ij} & = \frac{1}{N_{AB}}
 \begin{bmatrix}
  \displaystyle \tilde{S}^{-1}_{AA}\left(\sqrt{\frac{3}{f_{A}}}\right) -
  \frac{1}{f_{A}}\sqrt{\frac{3}{f_{A}}} &
  \displaystyle - \frac{1}{4 f_{A} f_{B}} \\
  \displaystyle - \frac{1}{4 f_{A} f_{B}} & 
  \displaystyle \tilde{S}^{-1}_{BB}\left(\sqrt{\frac{3}{f_{B}}}\right) -
  \frac{1}{f_{B}}\sqrt{\frac{3}{f_{B}}} \\
 \end{bmatrix}
\end{align}
where $\tilde{S}^{-1}(\xi)$ is the inverse matrix for the normalized scattering function
matrix $\tilde{S}(\xi)$ for the ideal systems.
\begin{equation}
 \sum_{j} \tilde{S}_{ij}(\xi) \tilde{S}^{-1}_{jk}(\xi) = \delta_{ik}
\end{equation}
$\tilde{S}(\xi)$ is defined as follows.
\begin{equation}
 \tilde{S}_{ij}(\xi) = 
  \begin{bmatrix}
   \displaystyle \frac{2}{\xi^{2}} (e^{- f_{A} \xi} - 1 + f_{A} \xi) & 
   \displaystyle \frac{1}{\xi^{2}} (e^{- f_{A} \xi} - 1)(e^{- f_{B} \xi} - 1) \\
   \displaystyle \frac{1}{\xi^{2}} (e^{- f_{A} \xi} - 1)(e^{- f_{B} \xi} - 1) &
   \displaystyle \frac{2}{\xi^{2}} (e^{- f_{B} \xi} - 1 + f_{B} \xi) \\
  \end{bmatrix}
\end{equation}

\section{Chemical Potential}
\label{chemical_potential}
The chemical potential for the system can be calculated by substituting
eq~\eqref{freeenergy_diblocksolution_psi} into
eq~\eqref{chemicalpotential}. The chemical potential for the A, B
subchains and S solvent is as follows.
\begin{equation}
 \begin{split}
  \mu_{A}(\bm{r}) = &
  \sum_{j (= A,B)}  \int d\bm{r}' \, 4 \sqrt{f_{A} f_{j}} A_{Aj}
  \mathcal{G}(\bm{r} - \bm{r}') \psi_{j}(\bm{r}') \\
  & + 2 f_{A} C_{AA} \left[ 2 \psi_{A}(\bm{r}) \ln \psi_{A}(\bm{r}) + \psi_{A}(\bm{r}) \right] \\
  & + 4 \sqrt{f_{A} f_{B}} C_{AB} \psi_{B}(\bm{r}) \\
  & - \frac{b^{2}}{3} \nabla^{2} \psi_{A}(\bm{r}) \\
  & + \sum_{j (= A,B,S)} 2 \chi_{A,j} \psi^{2}_{j}(\bm{r}) \psi_{A}(\bm{r})
 \end{split}
\end{equation}
\begin{equation}
 \begin{split}
  \mu_{B}(\bm{r}) = &
  \sum_{j (= A,B)}  \int d\bm{r}' \, 4 \sqrt{f_{B} f_{j}} A_{Bj}
  \mathcal{G}(\bm{r} - \bm{r}') \psi_{j}(\bm{r}') \\
  & + 2 f_{B} C_{BB} \left[ 2 \psi_{B}(\bm{r}) \ln \psi_{B}(\bm{r}) + \psi_{B}(\bm{r}) \right] \\
  & + 4 \sqrt{f_{A} f_{B}} C_{AB} \psi_{A}(\bm{r}) \\
  & - \frac{b^{2}}{3} \nabla^{2} \psi_{B}(\bm{r}) \\
  & + \sum_{j (= A,B,S)} 2 \chi_{B,j} \psi^{2}_{j}(\bm{r}) \psi_{B}(\bm{r})
 \end{split}
\end{equation}
\begin{equation}
 \begin{split}
  \mu_{S}(\bm{r}) = &
  \frac{2}{N_{S}} \left[ 2 \psi_{S}(\bm{r}) \ln \psi_{S}(\bm{r}) + \psi_{S}(\bm{r}) \right] \\
  & - \frac{b^{2}}{3} \nabla^{2} \psi_{S}(\bm{r}) \\
  & + \sum_{j (= A,B,S)} 2 \chi_{B,j} \psi^{2}_{j}(\bm{r}) \psi_{S}(\bm{r})
 \end{split}
\end{equation}
Note that $\mu_{i}(\bm{r})$ contains the Laplacian term $ -
b^{2} \nabla^{2} \psi_{i}(\bm{r}) / 3$ for all $i$.
Thus the evolution equation,
eq~\eqref{steepest_descent}, can be expressed as
\begin{equation}
 \label{steepest_descent_laplacian}
 \psi^{\text{(new)}}_{i}(\bm{r}) = \psi_{i}(\bm{r}) + \frac{\omega b^{2}}{3} \nabla^{2} \psi_{i}(\bm{r}) + \dots
\end{equation}
As mentioned before, eq~\eqref{steepest_descent_laplacian} can be solved stably by using
implicit scheme (the ADI scheme, in this work), in analogy to the diffusion equation.


%
%



%
%

\clearpage

\clearpage

\clearpage

\clearpage

\clearpage

\clearpage

\clearpage

\clearpage

\clearpage

\end{document}